\newif\iftwocol
\twocoltrue 

\iftwocol
\documentclass[journal,twocolumn,final]{IEEEtran}
\else
\documentclass[journal,onecolumn,draftcls,12pt]{IEEEtran}
\fi

\pdfoutput=1
\newlength{\figwidth}

\iftwocol
\else
\fi

\usepackage[nosort]{cite}        
\usepackage{url} 
\usepackage[intlimits]{amsmath}
\usepackage{bbm}
\usepackage{graphicx}
\usepackage{paralist}
\usepackage{fancyref}
\usepackage[stretch=16,shrink=16,step=4]{microtype}
\usepackage{vmr-symbols-rndbold}
\usepackage{standard-macros}
\safemath{\txant}{m\sub{t}} 
\safemath{\txantalt}{\widetilde{m}\sub{t}}
\safemath{\rxant}{m\sub{r}} 
\safemath{\cohtime}{n\sub{c}} 
\safemath{\bl}{n} 
\safemath{\err}{\epsilon} 
\let\snr\undefined
\safemath{\snr}{\rho} 
\safemath{\tfdiv}{l}  
\safemath{\ncod}{M}
\safemath{\realset}{\amsbb{R}}
\safemath{\naturalset}{\amsbb{N}}
\safemath{\Rmax}{R^*}
\safemath{\Rmaxala}{R^*\sub{ala}}
\safemath{\Rmaxp}{\Rmax(\tfdiv,\cohtime,\epsilon,\snr)}
\safemath{\Rmaxalap}{\Rmaxala(\tfdiv,\cohtime,\epsilon,\snr)}
\safemath{\Cerg}{C\sub{erg}}
\safemath{\Cout}{C\sub{out}}
\safemath{\Pout}{P\sub{out}}
\safemath{\diversity}{d}
\safemath{\multiplexing}{r}
\safemath{\Pustm}{P_{\rmatX}^{\mathrm{u}}}
\safemath{\maxantalt}{p}
\safemath{\minantalt}{q}
\safemath{\altgamma}{\widetilde{\gamma}}
\newcommand{\infoden}[2]{\ensuremath{\imath\lefto({#1};{#2}\right)}} 	
\safemath{\alshouffle}{e}

\makeatletter
\def\@IEEEinterspaceratioM{0.265}
\def\@IEEEinterspaceMINratioM{0.1651}
\def\@IEEEinterspaceMAXratioM{0.38}

\def\@IEEEinterspaceratioB{0.31}
\def\@IEEEinterspaceMINratioB{0.19}
\def\@IEEEinterspaceMAXratioB{0.38}
\@IEEEtunefonts
\makeatother
\let\markblue\undefined
\newcommand{\markblue}[1]{#1}

\begin{document}
\IEEEoverridecommandlockouts

%
\title{Short-Packet Communications over Multiple-Antenna Rayleigh-Fading Channels}
%
\author{Giuseppe~Durisi,~\IEEEmembership{Senior Member,~IEEE}, Tobias~Koch,~\IEEEmembership{Member,~IEEE}, Johan~\"Ostman,   Yury~Polyanskiy,~\IEEEmembership{Senior Member,~IEEE}, Wei~Yang,~\IEEEmembership{Member,~IEEE}
\thanks{This work was supported in part by the Swedish Research Council under grant 2012-4571, by the National Science Foundation CAREER award under grant  agreement CCF-12-53205, by the European Community's Seventh Framework Programme FP7/2007-2013 under Grant 333680, by the Ministerio de Econom\'ia y Competitividad of Spain under Grants RYC-2014-16322, TEC2013-41718-R, and CSD2008-00010, and by the Comunidad de Madrid under Grant S2013/ICE-2845. 
The simulations were performed in part on resources at Chalmers Centre for Computational Science and Engineering (C3SE) provided by the Swedish National Infrastructure for Computing (SNIC).}
\thanks{The material of this paper was presented in part at the 2012 IEEE Information Theory Workshop, Lausanne, Switzerland, and in part at the 2014 IEEE International Symposium on Wireless Communication Systems, Barcelona, Spain.}
\thanks{G. Durisi and J. \"Ostman are with the Department of Signals and Systems, Chalmers University of Technology, 41296,  Gothenburg, Sweden (e-mail: \{durisi,johanos\}@chalmers.se).}%
\thanks{T. Koch is with the Signal Theory and Communications Department, Universidad Carlos III de Madrid, 28911, Legan\'{e}s, Spain and with the Gregorio Mara\~n\'on Health Research Institute (e-mail: koch@tsc.uc3m.es).}
\thanks{Y. Polyanskiy is with the Department of Electrical Engineering and Computer
Science, MIT, Cambridge, MA, 02139 USA (e-mail: yp@mit.edu).}
\thanks{W. Yang is with the  Department of Electrical Engineering, Princeton University, NJ, 08544 USA (e-mail: weiy@princeton.edu) }
}
%

%
%
\maketitle

\begin{abstract}
Motivated by the current interest in ultra-reliable, low-latency, machine-type communication systems, we investigate the tradeoff between reliability, throughput, and latency in the transmission of information over multiple-antenna Rayleigh block-fading channels.
Specifically, we obtain finite-blocklength, finite-SNR upper and lower bounds on the maximum coding rate achievable over such channels for a given constraint on the packet error probability.
Numerical evidence suggests that our bounds delimit tightly the maximum coding rate already for short blocklengths (packets of about 100 symbols).
Furthermore, our bounds reveal the existence of a tradeoff between the rate gain obtainable by spreading each codeword over all available time-frequency-spatial degrees of freedom, and the rate loss caused by  the need of estimating the fading coefficients over these degrees of freedom.
In particular, our bounds allow us to determine the optimal number of transmit antennas and the optimal number of time-frequency diversity branches that maximize the rate.
Finally, we show that infinite-blocklength performance
metrics such as the ergodic capacity and the outage capacity yield inaccurate throughput estimates.
\end{abstract}


\section{Introduction}
Multi-antenna technology is a fundamental part of most modern wireless communication standards, due to its ability to provide tremendous gains in both spectral efficiency and reliability.  
The use of multiple antennas yields additional spatial degrees of freedom that can be used to lower the error probability for a given data rate, through the exploitation of \emph{spatial diversity}, or increase the data rate for a given error probability, through the exploitation of \emph{spatial multiplexing}.
These two effects cannot be harvested concurrently and there exists a fundamental tradeoff between diversity and multiplexing.
This tradeoff admits a particularly simple characterization in the high signal-to-noise ratio (SNR) regime~\cite{zheng03-05a}.

\markblue{Cellular systems offering mobile broadband services operate typically at maximum multiplexing~\cite{lozano10-09a} and do not make use of diversity-exploiting techniques such as space-time codes, whose purpose is to reduce the outage probability. 
Indeed, diversity-exploiting techniques are useless for low-mobility users, for which the fading coefficients can be learnt easily at the transmitter and outage events can be avoided altogether by rate adaptation.
They are not advantageous for high-mobility users as well, because of the abundant time and frequency selectivity that is available, which is sufficient for modern cellular systems to operate at the target outage level.}

These conclusions have been derived in~\cite{lozano10-09a} under the assumptions of long data packets ($1000$ channel uses or more) and moderately low packet-error rates (around $10^{-2}$), which are relevant for current mobile broadband services.

%

In next-generation (5G) cellular systems, it is expected that enhanced mobile-broadband services (exploiting most likely the millimeter-wave part of the frequency spectrum and relying on advanced antenna solutions) will be complemented by new services centered on \emph{machine-type communications} (MTC)~\cite{metis-project-deliverable-d1.113-04a,boccardi14-02a,popovski14-10a,fettweis14-02a,dahlman14-12a,durisi15-04a}.
An important emerging area among MTC systems is that of ultra-reliable, low-latency communications~\cite{johansson15-06a,yilmaz15-06a}, also known as \emph{mission-critical} MTC~\cite{dahlman14-12a}.
This area targets MTC systems that require reliable real-time communications with stringent requirements on latency, reliability, and availability. 
Examples of mission-critical MTCs include smart grids for power distribution automation, industrial manufacturing and control, and intelligent transportation systems~\cite{dahlman14-12a}. 
For example, in the case of industrial automation applications~\cite{johansson15-06a,yilmaz15-06a}, one is typically interested in transmitting short packets consisting of about $100$ bits within $100\,\mu\text{s}$  and with  $10^{-9}$ packet error rate. 

Motivated by mission-critical MTC systems, we investigate in this paper the fundamental tradeoff between throughput, reliability, and latency in short-packet wireless links.
\markblue{We also analyze how multiple antennas should be used in  such links.
Specifically, we address the following questions.
Can the stringent reliability requirements of mission-critical MTC be met if the available transmit antennas are used to increase throughput (i.e., provide spatial multiplexing), or should these antennas be used to increase reliability (i.e., provide spatial diversity)?}
What is the cost of learning the fading coefficients, whose knowledge is required to exploit the spatial degrees of freedom provided by multiple antennas, when the packet size is short?  Does this cost overcome the benefits of using multiple antennas?
%

\paragraph*{Contributions} 
\label{par:contributions}
The tension between reliability, throughput, and channel-estimation overhead in multiple-antenna communications have been investigated previously in the literature. 
However, as we shall review in Section~\ref{sec:maximum_coding_rate}, most of the available results are asymptotic either in the packet length~\cite{telatar99-11a,lozano10-09a,hassibi03-04a}, or in the SNR~\cite{zheng02-10a}, or in both~\cite{zheng02-02a,lapidoth02-05a,moser09-06a,yang13-02a}.
Hence, their relevance in the context of mission-critical MTC is unclear. 

In this paper, we address this issue by presenting a more refined \emph{nonasymptotic} analysis of the tradeoff between reliability, throughput, latency, and channel-estimation overhead, which relies on the finite-blocklength bounds developed in~\cite{polyanskiy10-05a}.
Our main contributions are as follows:
\begin{itemize}
  \item Focusing on the so-called \emph{Rayleigh block-fading} model~\cite{marzetta99-01a,zheng02-02a}, which is relevant for mission-critical MTC systems operating in a rich scattering environment~\cite{johansson15-06a,yilmaz15-06a}, we obtain nonasymptotic achievability and converse bounds on the maximum coding rate achievable for a given SNR,  a given packet size, and a given  packet reliability. 
  \item We present numerical evidence that the newly derived achievability and converse bounds delimit tightly the maximum coding rate  for packet lengths of interest  for mission-critical MTC systems.
  Furthermore, our numerical examples show that the bounds allow one to identify accurately the throughput-maximizing number of transmit antennas  as a function of the number of available time-frequency diversity branches.
  We also show that throughput estimates based on asymptotic performance metrics such as the ergodic capacity and the outage capacity are inaccurate, especially when the channel offers a significant amount of time-frequency diversity branches and the packet length is small.

  \item A comparison with nonasymptotic maximum coding rate bounds, obtained for specific space-time inner codes (such as the Alamouti scheme), allows us to identify when the available transmit antennas should be used to increase reliability, or throughput, or should be partly switched off to limit the channel-estimation overhead.
  %
\end{itemize}
In previous works, researchers have drawn inspiration from the structure of the capacity achieving distribution of multiple-input multiple-output (MIMO) channels to design practical coded-modulation schemes (see e.g.,~\cite{hochwald00-09a}). In this paper, we go one step further and study how the choice of the input distribution affects the nonasymptotic achievability bounds and the corresponding converse bounds.

The results in this paper generalize to the multiple-antenna setting the analysis conducted in~\cite{yang12-09a} for the single-input single-output case.
A partial extension of the results in~\cite{yang12-09a} to the MIMO case is provided in~\cite{ostman14-08b}. 
The analysis in~\cite{ostman14-08b}, however, relies critically on the assumption that the codewords are orthogonal in space and that the transmit power is allocated uniformly both across antennas and across coherence intervals (see~\cite[Eq.~(3)]{ostman14-08b}).  
This assumption is dropped in the current paper. 
As we shall illustrate in Section~\ref{sec:numerical_results}, allocating the power uniformly across antennas is in fact suboptimal when the number of available time-frequency diversity branches is large. 
Bounds on the maximum coding rate for the case of \emph{quasi-static} fading channels, i.e., channels for which the fading stays constant over the duration of each codeword are reported in~\cite{yang14-07c}.
Differently from~\cite{yang14-07c}, in this paper we allow each codeword to span multiple fading realizations in time and/or frequency.
%

\paragraph*{Notation} 
\label{par:notation}
Upper case letters such as $X$ denote scalar random variables and their realizations are written in lower case, e.g., $x$. We use boldface upper case letters to denote random vectors, e.g., $\rvecx$, and boldface lower case letters for their realizations, e.g., $\vecx$. Upper case letters of two special fonts are used to denote deterministic matrices (e.g., $\matY$) and random matrices (e.g., $\rmatY$). 
The superscripts~$^H$ and $^*$ stand for Hermitian transposition and complex conjugation, respectively, and we use $\tr\{\cdot\}$ and $\det\{\cdot\}$ to denote the trace and the determinant of a given matrix, respectively. 
The identity matrix of size $a\times a$ is written as~$\matI_a$. 
The distribution of a zero-mean, circularly symmetric complex Gaussian random variable with variance $\sigma^2$ is denoted by
$\jpg(0,\sigma^2)$.
For two functions $f(x)$ and $g(x)$, the notation $f(x) = \landauO(g(x))$, $x \to \infty$, means that
$\limsup_{x\to\infty} \abs{f(x)/g(x)}< \infty$, and $f(x) = \landauo(g(x))$, $x \to \infty$, means that $\lim_{x\to\infty}\abs{f(x)/g(x)} = 0$.
Finally, $\ln(\cdot)$ indicates the natural logarithm, $[a]^+ $ stands for $\max\lbrace a,0\rbrace$, and $\Gamma(\cdot)$ denotes the Gamma function.

Following~\cite{tse05a}, we say that a scheme provides time, frequency, or spatial diversity if it allows the information symbols to pass through independently fading signal paths (diversity branches) in time, frequency, or space.
We say that a scheme provides spatial multiplexing if it allows the transmission of multiple parallel data streams over the same channel.
Throughout the paper, we shall rely on these broad notions of diversity and multiplexing.
One exception is when we will review the diversity-multiplexing tradeoff (DMT)~\cite{zheng02-10a} in Section~\ref{sec:relation_to_previous_results}. 
To avoid any ambiguity, we shall refer to the quantities involved in the DMT, which are defined only in the high-SNR regime, as \emph{diversity gain} and \emph{multiplexing gain}.

\section{System Model} 
\label{sec:system_model}
We consider a Rayleigh block-fading channel with~$\txant$ transmit antennas and~$\rxant$ receive antennas that stays constant for $\cohtime$ channel uses.
For a frequency-flat narrowband channel, $\cohtime$ is the number of channel uses in time over which the channel stays constant (coherence time); for a frequency-selective channel and under the assumption that orthogonal frequency-division multiplexing (OFDM) is used, $\cohtime$ is the number of subcarriers over which the channel stays constant (coherence bandwidth). More generally,~$\cohtime$ can be interpreted as the number of ``time-frequency slots'' over which the channel does not change.

Within the $k$th coherence interval, the channel input-output relation can be written as
\begin{IEEEeqnarray}{rCl}
    \label{eq:channel_model}
    \rmatY_k = \matX_k \rmatH_k + \rmatW_k.
\end{IEEEeqnarray}
Here, $\matX_k \in \complexset^{\cohtime\times \txant}$ and $\rmatY_k \in \complexset^{\cohtime \times \rxant}$ are the transmitted and received matrices, respectively; the entries of the complex fading matrix $\rmatH_k \in \complexset^{\txant\times \rxant}$ are independent and identically distributed (i.i.d.) $\jpg(0,1)$; $\rmatW_k \in \complexset^{\cohtime \times \rxant}$ denotes the additive noise at the receiver and has i.i.d.\ $\jpg(0,1)$ entries.
We assume $ \left\lbrace \rmatH_k \right\rbrace$ and $ \left\lbrace \rmatW_k \right\rbrace$ to take on independent realizations over successive coherence intervals. 
We further assume that $\rmatH_k$ and $\rmatW_k$ are independent and that their joint law does not depend on $\matX_k$. 

Most throughput analyses available in the literature rely on the assumption that the receiver has perfect channel state information (CSI), i.e., that a ``genie'' informs the receiver about the realizations of the fading process~$\{\rmatH_k\}$.
As discussed in~\cite{lapidoth05-07a,moser09-06a,durisi11-08a},
this assumption relies on the fact that CSI can be acquired by transmitting some known training symbols that are used by the receiver to learn~$\{\rmatH_k\}$. 
Unfortunately, throughput estimates based on the assumption of perfect CSI at the receiver  are  overly optimistic for two reasons: 
\begin{inparaenum}[i)]
\item CSI will always be imperfect, no matter how long the training sequences are; 
\item transmitting training sequences yields a rate loss (channel-estimation overhead), which---as we shall see---can be significant for short-packet transmission. 
Analyses relying on the perfect-CSI assumption simply ignore this overhead.
\end{inparaenum}

To obtain more realistic throughput estimates, in this paper we drop  the assumption of perfect CSI at the receiver.
Instead, we assume that  the receiver has knowledge only of the statistics of the Rayleigh-fading process (i.e., its mean and its autocovariance function) but no \emph{a priori} knowledge of the realizations of $\{\rmatH_k\}$.
Note that this does not prevent the receiver from performing channel estimation. 
We merely view the transmission of training sequences to learn the channel at the receiver as a specific form of channel coding. This implies that in our setup the overhead associated with the transmission of such sequences is automatically accounted for.
%

Throughout the paper, we also assume no \emph{a priori} CSI at the transmitter. 
The transmitter has only knowledge of the statistics of the fading process.
This assumption is reasonable in a high-mobility scenario, where  fast channel variations make channel tracking at the transmitter unfeasible.
It is also appropriate for mission-critical applications where it may be desirable to avoid the creation of the feedback link required to provide CSI at the transmitter.



\section{Maximum Coding Rate} 
\label{sec:maximum_coding_rate}
We next introduce the notion of a channel code for the channel~\eqref{eq:channel_model}.
For simplicity, we shall restrict ourselves to codes whose blocklength $\bl$ is an integer multiple of the coherence interval $\cohtime$, i.e., $\bl=\tfdiv\cohtime$ for some $\tfdiv\in \naturalset$. 
\begin{dfn}\label{dfn:channel_code}
  An $(\tfdiv,\cohtime,\ncod,\epsilon,\snr)$ code for the channel~\eqref{eq:channel_model} consists of
  \begin{itemize}
    \item An encoder $f:\{1,\dots,\ncod\}\to\complexset^{\cohtime \times \txant\tfdiv}$ that maps the message $J\in\{1,\dots,\ncod\}$ to a codeword in the set $\{\matC_1,\dots,\matC_{\ncod}\}$.
    Since each codeword $\matC_m$, $m=1,\dots,\ncod$, spans $l$ coherence intervals, it is convenient to express it as the concatenation of $l$ subcodewords
    \begin{IEEEeqnarray}{rCL}
      \matC_m=\bigl[ \matC_{m,1}, \cdots, \matC_{m,l} \bigr].
    \end{IEEEeqnarray}
    We require that each subcodeword $\matC_{m,k}\in \complexset^{\cohtime\times\txant}$ satisfies the  power constraint
    \begin{IEEEeqnarray}{rCL}
      \label{eq:subavp}
      \tr\bigl\{\herm{\matC_{m,k}}\matC_{m,k}\bigr\}= \cohtime\snr,\,\, m=1,\dots,\ncod,\,\, k=1,\dots,\tfdiv. \IEEEeqnarraynumspace
    \end{IEEEeqnarray}
    Evidently,~\eqref{eq:subavp} implies the per-codeword  power constraint\footnote{It is more common in information-theoretic analyses to impose a power constraint per codeword and not per coherence interval. 
    The benefit of the per-codeword power constraint is that it leads to simple closed-form expressions
     for capacity. However, practical systems typically operate under
     constraint~\eqref{eq:subavp}.}
    %
    \begin{IEEEeqnarray}{rCL}
      \tr\bigl\{\herm{\matC_m}\matC_m\bigr\}&=& \tfdiv\cohtime\snr\label{eq:avp}\\
      &=&\bl \snr.
    \end{IEEEeqnarray}
    Since the noise has unit variance, $\snr$ in~\eqref{eq:avp} can be thought of as the SNR.

    \item A decoder $g:\complexset^{\cohtime\times\txant\tfdiv}\to \left\lbrace 1,\dots,\ncod\right\rbrace$ satisfying a maximum error probability constraint
    \begin{IEEEeqnarray}{rCL}
      \max_{1\leq j\leq \ncod} \Pr\lefto[g\bigl(\rmatY^\tfdiv\bigr)\neq J \given J=j\right]\leq \epsilon
    \end{IEEEeqnarray}
    where 
    \begin{IEEEeqnarray}{rCL}
      \rmatY^{\tfdiv}=\bigl[ \rmatY_{1}, \cdots,\rmatY_{\tfdiv} \bigr]
    \end{IEEEeqnarray}
    is the channel output induced by the transmitted codeword 
    \begin{IEEEeqnarray}{rCL}
      \matX^{\tfdiv}=\bigl[ \matX_{1},\cdots,\matX_{\tfdiv} \bigr]=f(j)
    \end{IEEEeqnarray}
     according to~\eqref{eq:channel_model}.
   \end{itemize}
\end{dfn}

The \emph{maximal channel coding rate} $\Rmaxp$ is defined as the largest rate ${(\ln \ncod)}/(\tfdiv\cohtime)$ for which there exists an $(\tfdiv,\cohtime,\ncod,\epsilon,\snr)$ code.
Formally,
\begin{IEEEeqnarray}{rCL}\label{eq:max_coding_rate}
  \Rmaxp\triangleq\sup\left\{\frac{\ln \ncod}{\tfdiv\cohtime} \sothat \exists (\tfdiv,\cohtime,\ncod,\epsilon,\snr) \text{ code} \right\}.\IEEEeqnarraynumspace
\end{IEEEeqnarray}
Recall that neither the encoder nor the decoder are assumed to have access to side information about the fading channel.
For the case when CSI is available at the receiver, $\Rmaxp$ has been characterized up to second order for specific scenarios in~\cite{polyanskiy11-08b,vituri12-11a,collins14-07a}.

\markblue{The maximal channel coding rate~\Rmaxp captures the fundamental tension between the  error probability $\epsilon$ and the transmission rate $R^*$ for a given blocklength $\bl=\tfdiv\cohtime$  and SNR $\snr$. 
Furthermore, its dependency on the coherence interval $n_c$, on the number of diversity branches $\tfdiv$, and on the number of transmit and receive antennas\footnote{\markblue{This dependency is not made explicit in the notation used in~\eqref{eq:max_coding_rate}, in order to keep the notation compact.}} allows one to study how this tension depends on the characteristics of the fading channel.}

\section{Relation to Previous Results} 
\label{sec:relation_to_previous_results}
Most of the results available in the literature can be interpreted as asymptotic characterizations of $\Rmaxp$ for  $\tfdiv\to \infty$, or $\cohtime\to \infty$, or $\snr\to\infty$, or a combination of these limits.
\paragraph*{Ergodic capacity} 
\label{par:ergodic_capacity}
For the case when $\tfdiv\to \infty$ for fixed $\cohtime$, fixed $\snr$, and fixed $0<\epsilon<1$, the maximum coding rate $\Rmaxp$ converges to the ergodic capacity $\Cerg(\snr)$
\begin{IEEEeqnarray}{rCL}\label{eq:erg_cap}
  \lim_{\tfdiv \to \infty} \Rmaxp = \Cerg(\snr)=\frac{1}{\cohtime}\sup I(\rmatX;\rmatY)
\end{IEEEeqnarray}
where $\rmatX \in \complexset^{\cohtime\times \txant}$ denotes the channel input, $\rmatY\in\complexset^{\cohtime\times\rxant}$ is the corresponding channel output, obtained through~\eqref{eq:channel_model}, and the supremum in~\eqref{eq:erg_cap} is over all probability distributions on $\rmatX$ satisfying $\tr\{\herm{\rmatX}\rmatX\}=\cohtime\snr$ almost surely.
Note that, by the strong converse~\cite{wolfowitz57-12a}, the ergodic capacity $\Cerg(\snr)$ does not depend on $\epsilon$.
Although $\Cerg(\snr)$ is not known in closed form when  CSI is not available \emph{a priori} at the receiver, its high-SNR behavior is well understood~\cite{marzetta99-01a,hochwald00-03a,hassibi02-06a,zheng02-02a,yang13-02a}.
Specifically, Zheng and Tse~\cite{zheng02-02a} showed that, under the assumption $\cohtime>1$, 
\begin{IEEEeqnarray}{rCL}\label{eq:erg_high_snr}
  \Cerg(\snr)=m^* \left(1-\frac{m^*}{\cohtime}\right)\ln \snr + \landauO(1), \quad \snr\to \infty
\end{IEEEeqnarray}
where 
\begin{IEEEeqnarray}{rCL}\label{eq:txant_opt}
   m^*=\min\{\txant,\rxant,\floor{\cohtime/2}\}. 
\end{IEEEeqnarray}
%
%
%
We remark that~\eqref{eq:erg_high_snr} holds also when the maximization in~\eqref{eq:erg_cap} is performed under the less stringent constraint that $\Ex{}{\tr\{\herm{\rmatX}\rmatX\}}\leq \cohtime\snr$.
Since $\Cerg(\snr)=\min\{\txant,\rxant\}\ln \snr+\landauO(1)$ for the case when the receiver has perfect CSI~\cite{telatar99-11a}, we see from~\eqref{eq:erg_high_snr} that the \emph{prelog} penalty due to lack of \emph{a priori} CSI is equal to $(m^*)^2/\cohtime$ (provided that $\cohtime\geq \txant+\rxant$). 
This is roughly $m^*$ times the number of pilots per time-frequency slot needed to learn the channel at the receiver when $m^*$ transmit antennas are used.
The prelog penalty vanishes as $\cohtime$ becomes large.

By tightening the high-SNR expansion~\eqref{eq:erg_high_snr}~\cite{zheng02-02a,yang13-02a}, one obtains an accurate finite-SNR approximation of capacity~\cite{yang12-09a,devassy15-07a}.
The input distribution that achieves the first two terms in the resulting high-SNR expansion of $\Cerg(\snr)$ depends on the relationship between $\cohtime$, $\txant$ and $\rxant$.
When $\cohtime\geq \txant+\rxant$, it is optimal at  high SNR to choose $\rmatX$ to be a scaled  isotropically distributed matrix that has orthonormal columns~\cite{zheng02-02a}.
This input distribution is sometimes referred to as USTM.
When $\cohtime<\txant+\rxant$, Beta-variate space-time modulation (BSTM) should be used instead~\cite{yang13-02a}.
In BSTM, the USTM unitary matrix is multiplied by a diagonal matrix whose nonzero entries are distributed as the square-root of the eigenvalues of a Beta-distributed random matrix.
Throughout this paper, we shall focus on the case $\cohtime\geq \txant+\rxant$.

Although the ergodic capacity captures the rate penalty due to the channel-estimation overhead, and although its high-SNR expansion~\eqref{eq:erg_high_snr} describes compactly how this penalty depends on the channel coherence interval, the infinite-blocklength nature of~\eqref{eq:erg_cap} and its independence on the packet reliability $\epsilon$ limit its usefulness for the short-packet scenario considered in this paper.
%
%
\paragraph*{Outage capacity} 
\label{par:outage_capacity}
%
For the case when $\cohtime\to\infty$ for fixed $\tfdiv$, $\epsilon$, and $\snr$, the maximum coding rate $\Rmaxp$ converges to the outage capacity $\Cout(\snr,\epsilon)$, defined as~\cite{ozarow94-05a}
\iftwocol
\begin{multline}
  \lim_{\cohtime\to\infty}\Rmaxp=\Cout(\snr,\epsilon) \\
   =\sup\left\{R \sothat \inf_{\{\matQ_k\}_{k=1}^{\tfdiv}} \Pout\lefto(\{\matQ_k\}_{k=1}^\tfdiv,R\right)\leq \epsilon \right\}.\label{eq:outage_capacity}
\end{multline}
\else
\begin{IEEEeqnarray}{rCL}
 \lim_{\cohtime\to\infty}\Rmaxp&=&\Cout(\snr,\epsilon) \notag\\
  &=&\sup\left\{R \sothat \inf_{\{\matQ_k\}_{k=1}^{\tfdiv}} \Pout\lefto(\{\matQ_k\}_{k=1}^\tfdiv,R\right)\leq \epsilon \right\}.\IEEEeqnarraynumspace\label{eq:outage_capacity}
\end{IEEEeqnarray}
\fi
Here, $\Pout(\cdot,\cdot)$ is the outage probability
\iftwocol
\begin{multline}
   \Pout\lefto(\{\matQ_k\}_{k=1}^{\tfdiv},R\right)\\=\Pr\lefto\{\frac{1}{l}\sum_{k=1}^{l} \ln\det(\matI_{\rxant}+\herm{\rmatH}_k\matQ_k\rmatH_k)\leq R\right\}
\end{multline}
\else
\begin{IEEEeqnarray}{rCL}\label{eq:outage_probability}
  \Pout\lefto(\{\matQ_k\}_{k=1}^{\tfdiv},R\right)=\Pr\lefto\{\frac{1}{l}\sum_{k=1}^{l} \ln\det(\matI_{\rxant}+\herm{\rmatH}_k\matQ_k\rmatH_k)\leq R\right\} \IEEEeqnarraynumspace
\end{IEEEeqnarray}
\fi
where, for the Rayleigh-fading case considered in this paper, $\{\matQ_k\}$, $k=1,\dots,\tfdiv$, are $\txant\times\txant$ diagonal matrices with nonnegative entries that satisfy $\tr\{\matQ_k\}=\snr$, and where the infimum in~\eqref{eq:outage_capacity} is over all~$\{\matQ_k\}$.
For the case $\tfdiv=1$, Telatar~\cite{telatar99-11a}   conjectured that the optimal diagonal matrix $\matQ_1$ is of the form
\begin{IEEEeqnarray}{rCL}
  \matQ_1=\frac{\snr}{m}\diag\{\underbrace{1,\dots,1}_{m},\underbrace{0,\dots,0}_{\txant-m}\}
\end{IEEEeqnarray}
for some $m \in \{1,\dots,\txant\}$.
This conjecture was proved in~\cite{abbe13-05a} for the multiple-input single-output case.

The outage capacity in~\eqref{eq:outage_capacity} characterizes in an implicit way the tension between the reliability $\epsilon$ and the throughput $R$.
Note that~\eqref{eq:outage_capacity} holds irrespectively of whether CSI is available at the receiver or not.
Indeed, as the coherence interval~$\cohtime$ gets large, the cost of learning the channel at the receiver vanishes~\cite[p.~2632]{biglieri98-10a},\cite{yang14-07c}.
Consequently, analyses based on outage capacity do not capture the overhead due to channel estimation, which may be significant for short-packet communications.

\paragraph*{Diversity-multiplexing tradeoff} 
\label{par:diversity_multiplexing_tradeoff}
Consider the scenario where $\tfdiv$ and $\cohtime$ are fixed, CSI is available at the receiver, and the packet error rate $\epsilon$ vanishes as a function of $\snr$ according to
%
%
\begin{IEEEeqnarray}{rCL}\label{eq:diversity_gain}
  \epsilon(\snr)=\snr^{-\diversity\,\tfdiv}
\end{IEEEeqnarray}
where $\diversity\in\{0,1,\dots,\txant\rxant\}$ is the so-called \emph{spatial diversity gain}. 
For the case when $\cohtime\geq \txant+\rxant-1$, Zheng and Tse proved that~\cite{zheng03-05a}
\begin{IEEEeqnarray}{rCL}\label{eq:dmt_formulation}
  \lim_{\snr\to\infty}\frac{\Rmax(\cohtime,\tfdiv,\epsilon(\snr),\snr)}{\ln\snr}=\multiplexing(\diversity)
\end{IEEEeqnarray}
where the \emph{multiplexing gain} $\multiplexing(\diversity)$ is the piece-wise linear function connecting the points
\begin{IEEEeqnarray}{rCL}\label{eq:dmt_coherent}
  \multiplexing\bigl((\txant-k)(\rxant-k)\bigr)=k, \quad k=0,\dots,\min\{\txant,\rxant\}.\IEEEeqnarraynumspace
\end{IEEEeqnarray}
The condition $\cohtime\geq \txant+\rxant-1$ has been relaxed to $\cohtime\geq \txant$ in~\cite{elia06-09a}, where an explicit  code construction that achieves~\eqref{eq:dmt_formulation} is provided.

For the case when CSI is not available at the receiver and $\cohtime\geq 2m^*+\rxant+1$ (where $m^*$ is given in~\eqref{eq:txant_opt}), the diversity-multiplexing tradeoff becomes~\cite{zheng02-10a,zheng02-11a}
\begin{IEEEeqnarray}{rCL}\label{eq:dmt_noncoherent} \lim_{\snr\to\infty}\frac{\Rmax(\cohtime,\tfdiv,\epsilon(\snr),\snr)}{\ln\snr}=\left(1-\frac{m^*}{\cohtime}\right)\multiplexing(\diversity).
\end{IEEEeqnarray}
The expressions in~\eqref{eq:dmt_coherent} and in~\eqref{eq:dmt_noncoherent} describe elegantly and succinctly the tradeoff between diversity gain and multiplexing gain.
The price to be paid for such a characterization is its high-SNR nature, which may limit its significance for the scenarios analyzed in this paper. 

Finite-SNR versions of the DMT have been proposed in~\cite{azarian07-02a,loyka10-10a}. 
However, these extensions rely on the outage probability and are, in contrast to the original formulation in~\cite{zheng02-10a}, only meaningful asymptotically as the blocklength tends to infinity.

To summarize, the performance metrics developed so far for the analysis of wireless systems, i.e., the ergodic capacity, the outage capacity, and the DMT have shortcomings when applied to short-packet wireless communications.
We address these shortcomings in the next section by developing nonasymptotic bounds on $\Rmaxp$.

\section{Bounds on the Maximal Coding Rate} 
\label{sec:bounds_on_the_maximal_coding_rate}
\subsection{Output Distribution Induced by USTM Inputs} 
\label{sec:ustm_output_distribution_in_closed_form}
Let $\rmatA$ be an $n\times m$ ($n>m$) random matrix.
We say that $\rmatA$ is isotropically distributed if, for every deterministic $n\times n$ unitary matrix $\matV$, the matrix $\matV\rmatA$ has the same probability distribution as~$\rmatA$.
A key ingredient of the nonasymptotic bounds on $\Rmaxp$ described in this section is the following closed-form expression for the probability density function (pdf) induced on the channel output $\rmatY_k$ in~\eqref{eq:channel_model} when $\rmatX_k$ is a scaled isotropically distributed matrix with orthonormal columns. 
Such an input distribution is commonly referred to as USTM.
It will turn out convenient to consider a minor modification of the USTM distribution, in which only $\txantalt$ out of the available $\txant$ transmit antennas are used. 
\begin{lem}\label{lem:ustm_output}
  Assume that $\cohtime\geq \txant+\rxant$. 
  Let $\minantalt=\min\{\txantalt,\rxant\}$ and $\maxantalt=\max\{\txantalt,\rxant\}$.
  Let also $\rmatX=\sqrt{\snr\cohtime/\txantalt}\rmatU$ where  $\rmatU\in\complexset^{\cohtime\times\txantalt}$ ($1\leq \txantalt\leq \txant$) satisfies $\herm{\rmatU}\rmatU=\matI_{\txantalt}$ and is isotropically distributed. 
Further, let  $\rmatY=\rmatX\rmatH+\rmatW$ where $\rmatH\in \complexset^{\txantalt\times\rxant}$ and $\rmatW \in \complexset^{\cohtime\times \rxant}$ are defined as in~\eqref{eq:channel_model}.
  The pdf of $\rmatY$ is given by
  \iftwocol
  \begin{IEEEeqnarray}{rCL}
   f_{\rmatY}(\matY)&=& \frac{\prod\limits_{u=\cohtime-\minantalt+1}^{\cohtime}\Gamma(u)}{\pi^{\rxant\cohtime}\prod\limits_{u=1}^{\txantalt}\Gamma(u)}
   \frac{(1+\mu)^{\txantalt(\cohtime-\txantalt-\rxant)}}{\mu^{\txantalt(\cohtime-\txantalt)}}\IEEEeqnarraynumspace\notag\\
   &&\cdot 
   \psi_{\txantalt}(\sigma_1^2,\dots,\sigma_{\rxant}^2).\label{eq:USTM_output_distribution}
  \end{IEEEeqnarray}
  \else
  \begin{IEEEeqnarray}{rCL}
   f_{\rmatY}(\matY)&=& \frac{\prod\limits_{u=\cohtime-\minantalt+1}^{\cohtime}\Gamma(u)}{\pi^{\rxant\cohtime}\prod\limits_{u=1}^{\txantalt}\Gamma(u)}
   \frac{(1+\mu)^{\txantalt(\cohtime-\txantalt-\rxant)}}{\mu^{\txantalt(\cohtime-\txantalt)}}
   %
   \psi_{\txantalt}(\sigma_1^2,\dots,\sigma_{\rxant}^2).\label{eq:USTM_output_distribution}
  \end{IEEEeqnarray}
  \fi
  Here, $\sigma_1>\dots>\sigma_{\rxant}$ denote the $\rxant$ nonzero singular values of $\matY$, which are positive and distinct almost surely~\cite{tulino04a}, $\mu=\snr\cohtime/\txantalt$, and
  \begin{IEEEeqnarray}{rCL}\label{eq:def_psi_function}    
    \psi_{\txantalt}(\sigma^2_1,\dots,\sigma^2_{\rxant})&=&\frac{\det\{\matM\}}{\prod\limits_{i<j}^{\rxant}(\sigma^2_i-\sigma^2_j)} 
    %
    \prod\limits_{k=1}^{\rxant}\frac{e^{-\sigma^2_k/(1+\mu)}}{\sigma_k^{2(\cohtime-\rxant)}}. \IEEEeqnarraynumspace
  \end{IEEEeqnarray}
 The entries of the $\maxantalt\times\maxantalt$ real matrix $\matM$   are given by
  \iftwocol
  \begin{IEEEeqnarray}{rCL}
    [\matM]_{ij}=
    \begin{cases}
      b_i^{\txantalt-j}\altgamma\bigl(\cohtime+j-\maxantalt-\txantalt,b_i\mu/(1+\mu)\bigr), \\
      \quad\quad 1\leq i\leq \rxant, \quad  1\leq j\leq \txantalt\\[1mm]
      \displaystyle e^{-b_i\mu/(1+\mu)}\, \left[\frac{\partial^{\txantalt-j}}{\partial \delta^{\txantalt-j}} \delta^{\cohtime-i} \bigg\rvert_{\delta=\frac{\mu}{1+\mu}} \right],\IEEEeqnarraynumspace\\
      \quad\quad \rxant < i \leq \maxantalt, \quad 1\leq j \leq \txantalt\\[1mm]
       b_i^{\cohtime-j} e^{-b_i \mu/(1+\mu)}\\
       \quad \quad  1\leq i\leq \rxant, \quad \txantalt< j\leq \maxantalt
    \end{cases}
  \end{IEEEeqnarray}
  %
  \else
  \begin{IEEEeqnarray}{rCL}
    [\matM]_{ij}=
    \begin{cases}
      \sigma_i^{2(\txantalt-j)}\altgamma\bigl(\cohtime+j-\maxantalt-\txantalt,\sigma^2_i\mu/(1+\mu)\bigr), & 1\leq i\leq \rxant \\
      &  1\leq j\leq \txantalt\IEEEeqnarraynumspace\\[1mm]
      \displaystyle \exp\bigl(-\sigma^2_i\mu/(1+\mu)\bigr)\, \left[\frac{\partial^{\txantalt-j}}{\partial \alpha^{\txantalt-j}} \alpha^{\cohtime-i} \bigg\rvert_{\alpha=\mu/(1+\mu)} \right], & \rxant < i \leq \maxantalt\\
      & 1\leq j \leq \txantalt\\[1mm]
       \sigma_i^{2(\cohtime-j)} \exp\bigl(-\sigma^2_i \mu/(1+\mu)\bigr), & 1\leq i\leq \rxant \\
      & \txantalt< j\leq \maxantalt
    \end{cases}
  \end{IEEEeqnarray}
  \fi
  where $b_i=\sigma_i^2$, $i=1,\dots,\rxant$, and
  \begin{IEEEeqnarray}{rCL}
    \altgamma(n,x) \triangleq \frac{1}{\Gamma(n)}\int\nolimits_{0}^{x}t^{n-1} e^{-t} dt
  \end{IEEEeqnarray}
  denotes the regularized incomplete Gamma function.
\end{lem}
\begin{IEEEproof}
  The proof, which relies on the Itzykson-Zuber integral~\cite[Eq.~(3.2)]{itzykson80-a} and on repeated use of~\cite[Lem.~5]{ghaderipoor12-05a}, can be found, e.g., in~\cite[App.~A]{yang13-02a} and, more recently, in~\cite{alfano14-10a}.
\end{IEEEproof}

\begin{rem}
A different expression for $f_{\rmatY}(\matY)$ can be found in~\cite{hassibi02-06a}. 
The expression in Lemma~\ref{lem:ustm_output}  appears to be easier to compute and more stable numerically.
\end{rem}

\subsection{USTM Dependence-Testing (DT) Lower Bound} 
\label{sec:ustm_dependence_testing_lower_bound}
We first present a lower bound on \Rmaxp that is based on the  dependence-testing (DT) bound~\cite[Th.~22]{polyanskiy10-05a} (maximal error probability) and makes use of the USTM-induced output distribution given in Lemma~\ref{lem:ustm_output}.
\begin{thm}\label{thm:dt_lb}
  Let  $\Lambda_{k,\txantalt,1}>\dots>\Lambda_{k,\txantalt,\rxant}$ be the ordered eigenvalues of  $\herm{\rmatZ}_{k}\matD_{\txantalt}\rmatZ_{k}$
  where $\{\rmatZ_{k} \}_{k=1}^{\tfdiv}$ are independent complex Gaussian  $\cohtime\times \rxant$ matrices with \iid $\jpg(0,1)$ entries, and 
  \begin{IEEEeqnarray}{rCL}  \matD_{\txantalt}=\diag\biggl\{\underbrace{1+\snr\cohtime/\txantalt,\dots,1+\snr\cohtime/\txantalt}_{\txantalt}, \underbrace{1, \dots, 1}_{\cohtime-\txantalt}\biggr\}\IEEEeqnarraynumspace
  \end{IEEEeqnarray}
  for $\txantalt \in \{1,\dots,\txant\}$.
  It can be shown that the eigenvalues are positive and distinct almost surely.
  %
 Let
  \iftwocol
  \begin{IEEEeqnarray}{rCL}\label{eq:S} 
    S_{k,\txantalt}&=&\txantalt(\cohtime-\txantalt)\ln\frac{\snr\cohtime}{\txantalt+\snr\cohtime}-\sum_{u=\cohtime-\minantalt+1}^{\cohtime} \ln\Gamma(u)
        \notag\\
        &&
        +\sum_{u=1}^{\txantalt}\ln\Gamma(u)
        -\tr\bigl\{\herm{\rmatZ}_{k}\rmatZ_{k}\bigr\}\notag\\
    &&
    -\ln\psi_{\txantalt}(\Lambda_{k,\txantalt,1},\dots,\Lambda_{k,\txantalt,\rxant})
  \end{IEEEeqnarray}
  \else
  \begin{IEEEeqnarray}{rCL}\label{eq:S} 
    S_{k,\txantalt}&=&\txantalt(\cohtime-\txantalt)\ln\frac{\snr\cohtime}{\txantalt+\snr\cohtime}-\sum_{u=\cohtime-\minantalt+1}^{\cohtime} \ln\Gamma(u)
        +\sum_{u=1}^{\txantalt}\ln\Gamma(u)\notag\\
        &&-\tr\bigl\{\herm{\rmatZ}_{k}\rmatZ_{k}\bigr\}
    -\ln\psi_{\txantalt}(\Lambda_{k,\txantalt,1},\dots,\Lambda_{k,\txantalt,\rxant})
  \end{IEEEeqnarray}
  \fi
  where $\minantalt=\min\{\txantalt,\rxant\}$ and the function $\psi_{\txantalt}:\positivereals^{\rxant}\to\reals$ was defined in~\eqref{eq:def_psi_function}.
  Finally, let
  %
  %
  %
  %
  %
%
%
\iftwocol
\begin{IEEEeqnarray}{rCL}\label{eq:error_rate_DT}
  \epsilon\sub{ub}(\ncod)=\min_{1\leq \txantalt\leq \txant}   \Ex{}{e^{-\left[\sum_{k=1}^{\tfdiv}S_{k,\txantalt}-\ln(\ncod-1)\right]^{+}}}.
\end{IEEEeqnarray}
\else
\begin{IEEEeqnarray}{rCL}\label{eq:error_rate_DT}
  \epsilon\sub{ub}(\ncod)=\min_{1\leq \txantalt\leq \txant}   \Ex{}{\exp\lefto\{-\left[\sum_{k=1}^{\tfdiv}S_{k,\txantalt}-\ln(\ncod-1)\right]^{+}\right\}}.
\end{IEEEeqnarray}
\fi
  Then
  \begin{IEEEeqnarray}{rCL}\label{eq:dt_bound}
    \Rmaxp\geq \max \left\{ \frac{\ln \ncod}{\cohtime\tfdiv} \sothat  \epsilon\sub{ub}(\ncod)\leq \epsilon\right\}.
  \end{IEEEeqnarray}
\end{thm}
\begin{IEEEproof}
See Appendix~\ref{sec:proof_of_theorem_dt}.
\end{IEEEproof}

\subsection{Meta-converse (MC) Upper Bound} 
\label{sec:metaconverse_upper_bound}
We next give an upper bound on $\Rmaxp$ that is based on the meta-converse (MC) theorem for maximal error probability of error~\cite[Th.~31]{polyanskiy10-05a} and uses the output distribution induced by the USTM input distribution (see~\eqref{eq:USTM_output_distribution}) as auxiliary output distribution.
\begin{thm}\label{thm:metaconverse_upper_bound}
    For a fixed $\txantalt\in [1,\dots,\txant]$, let the random variables $\{\bar{\rmatY}_k\}_{k=1}^{\tfdiv}$ be \iid $f_{\rmatY}$-distributed, with $f_{\rmatY}$, defined in~\eqref{eq:USTM_output_distribution}, being the output distribution corresponding to an USTM input distribution over $\txantalt$ antennas.
    Let $\Delta_{k,\txantalt,1}>\,\cdots\, > \Delta_{k,\txantalt,\rxant}$ be the ordered eigenvalues of $\herm{\bar{\rmatY}_k}\bar{\rmatY}_k$, $k=1,\dots,\tfdiv$,  and let
    \begin{IEEEeqnarray}{rCL}
        \rmatDelta_{k,\txantalt}=\diag\{\Delta_{k,\txantalt,1},\dots,\Delta_{k,\txantalt,\rxant}\}.
    \end{IEEEeqnarray}
    %
%
It can be shown that the eigenvalues are positive and distinct almost surely.
Let $\{\matSigma_k\}_{k=1}^{\tfdiv}$ be $\txant\times\txant$ \emph{diagonal} matrices with nonnegative diagonal entries, satisfying $\tr\{\matSigma_k\}= \cohtime\snr$, $k=1,\dots,\tfdiv$.
Let
\begin{IEEEeqnarray}{rCL}
  \widetilde{\matSigma}_k=\mat
                          \matI_{\txant}+\matSigma_k & \veczero\\
                          \veczero  & \matI_{\cohtime-\txant}
                          \emat.
\end{IEEEeqnarray}
Further let $\{\rmatU_k\}_{k=1}^{\tfdiv}$ be \iid isotropically distributed (truncated) $\cohtime\times\rxant$ unitary matrices, and let $\{\bar{\rmatZ}_k\}_{k=1}^{\tfdiv}$ be independent complex Gaussian $\cohtime\times\rxant$ matrices with \iid $\jpg(0,1)$ entries.
Finally,
let
\iftwocol
\begin{IEEEeqnarray}{rCL}
  &\bar{c}_{\txantalt}&(\matSigma_k)=\txantalt(\cohtime-\txantalt)\ln\frac{\snr\cohtime}{\txantalt}\notag\\
  &&
  -\txantalt(\cohtime-\txantalt-\rxant)\ln\lefto(1+\frac{\snr\cohtime}{\txantalt}\right) \notag \\
&&-\rxant\ln\det\widetilde{\matSigma}_k-\sum_{u=\cohtime-p+1}^{\cohtime} \ln\Gamma(u) +\sum_{u=1}^{\txantalt}\ln\Gamma(u)
\IEEEeqnarraynumspace
\end{IEEEeqnarray}
\else
\begin{IEEEeqnarray}{rCL}
  \bar{c}_{\txantalt}(\matSigma_k)&=& \txantalt(\cohtime-\txantalt)\ln\frac{\snr\cohtime}{\txantalt}
  -\txantalt(\cohtime-\txantalt-\rxant)\ln\lefto(1+\frac{\snr\cohtime}{\txantalt}\right) \notag \\
&&-\rxant\ln\det\widetilde{\matSigma}_k-\sum_{u=\cohtime-p+1}^{\cohtime} \ln\Gamma(u) +\sum_{u=1}^{\txantalt}\ln\Gamma(u)
\IEEEeqnarraynumspace
\end{IEEEeqnarray}
\fi
\iftwocol
\begin{IEEEeqnarray}{rCL}\label{eq:Q_inf}
  T_{k,\txantalt}(\matSigma_k)&=&\bar{c}_{\txantalt}(\matSigma_k) -\tr\{\rmatU_k\rmatDelta_{k,\txantalt}\herm{\rmatU_k}\widetilde{\matSigma}_k^{-1})\} \notag \\
  &&
  -\ln \psi_{\txantalt}(\Delta_{k,\txantalt,1},\dots,\Delta_{k,\txantalt,\rxant})
\end{IEEEeqnarray}
\else
\begin{IEEEeqnarray}{rCL}\label{eq:Q_inf}
  T_{k,\txantalt}(\matSigma_k)&=&\bar{c}_{\txantalt}(\matSigma_k) -\tr\{\rmatU_k\rmatDelta_{k,\txantalt}\herm{\rmatU_k}\widetilde{\matSigma}_k^{-1})\} 
  %
  -\ln \psi_{\txantalt}(\Delta_{k,\txantalt,1},\dots,\Delta_{k,\txantalt,\rxant})
\end{IEEEeqnarray}
\fi
and
\iftwocol
\begin{IEEEeqnarray}{rCL}\label{eq:P_inf}
  \bar{S}_{k,\txantalt}(\matSigma_k)&=& \bar{c}_{\txantalt}(\matSigma_k)-\tr\{\herm{\bar{\rmatZ}}_k\bar{\rmatZ}_k\} \notag\\
  &&
  -\ln \psi_{\txantalt}(\bar{\Lambda}_{k,\txantalt,1},\dots,\bar{\Lambda}_{k,\txantalt,\rxant}).
\end{IEEEeqnarray}
\else
\begin{IEEEeqnarray}{rCL}\label{eq:P_inf}
  \bar{S}_{k,\txantalt}(\matSigma_k)&=& \bar{c}_{\txantalt}(\matSigma_k)-\tr\{\herm{\bar{\rmatZ}}_k\bar{\rmatZ}_k\} 
  -\ln \psi_{\txantalt}(\bar{\Lambda}_{k,\txantalt,1},\dots,\bar{\Lambda}_{k,\txantalt,\rxant}).
\end{IEEEeqnarray}
\fi
Here, $\bar{\Lambda}_{k,\txantalt,1}>\,\cdots>\,\bar{\Lambda}_{k,\txantalt,\rxant}$ are the ordered  eigenvalues of $\herm{\rmatZ_k}\widetilde{\matSigma}_k\rmatZ_k$ (which are positive and distinct almost surely), $\maxantalt=\max\{\txantalt,\rxant\}$, and $\psi_{\txantalt}$ is defined in~\eqref{eq:def_psi_function}.
Then, for every $\bl$ and for every $0<\epsilon<1$, the maximal channel coding rate \Rmaxp is upper bounded by
\iftwocol
\begin{multline}\label{eq:metaconverse_bound}
  \Rmaxp\leq\\
   \min_{1\leq \txantalt\leq \txant} \sup_{\{\matSigma_k\}_{k=1}^{\tfdiv}} \frac{1}{n} \ln \frac{1}{\displaystyle\Pr\lefto\{\sum_{k=1}^{\tfdiv}T_{k,\txantalt}(\matSigma_k)\geq \gamma\right\}}
\end{multline}
\else
\begin{IEEEeqnarray}{rCL}\label{eq:metaconverse_bound}
  \Rmaxp\leq
   \min_{1\leq \txantalt\leq \txant} \sup_{\{\matSigma_k\}_{k=1}^{\tfdiv}} \frac{1}{n} \ln \frac{1}{\displaystyle\Pr\lefto\{\sum_{k=1}^{\tfdiv}T_{k,\txantalt}(\matSigma_k)\geq \gamma\right\}}
\end{IEEEeqnarray}
\fi
where
$\gamma=\gamma\lefto(\{\matSigma_k\}_{k=1}^{\tfdiv}\right)$ is the solution of
\begin{IEEEeqnarray}{rCL}
  \Pr\lefto\{\sum_{k=1}^{\tfdiv}\bar{S}_{k,\txantalt}(\matSigma_k)\leq \gamma\right\}=\epsilon.
\end{IEEEeqnarray}
\end{thm} 
\begin{IEEEproof}
  See Appendix~\ref{sec:proof_of_theorem_MC}.
\end{IEEEproof}

\begin{rem}
  To facilitate its numerical evaluation, the MC upper bound~\eqref{eq:metaconverse_bound} can be relaxed by using~\cite[Eq.~(102)]{polyanskiy10-05a}, which yields 
\iftwocol
\begin{multline}\label{eq:information_spectrum}
  \Rmaxp\leq \min_{1\leq \txantalt\leq \txant} \sup_{\{\matSigma_k\}_{k=1}^{\tfdiv}} \inf_{\lambda>0} \\
   \frac{1}{n} \lefto[\lambda-\ln\lefto( \left[\Pr\lefto\{\sum_{k=1}^{\tfdiv}\bar{S}_{k,\txantalt}(\matSigma_k)\leq \lambda\right\} -\epsilon\right]^{+}\right)\right].
\end{multline}
\else
\begin{IEEEeqnarray}{rCL}\label{eq:information_spectrum}
  \Rmaxp\leq \min_{1\leq \txantalt\leq \txant} \sup_{\{\matSigma_k\}_{k=1}^{\tfdiv}} \inf_{\lambda>0} 
   \frac{1}{n} \lefto[\lambda-\ln\lefto( \left[\Pr\lefto\{\sum_{k=1}^{\tfdiv}\bar{S}_{k,\txantalt}(\matSigma_k)\leq \lambda\right\} -\epsilon\right]^{+}\right)\right].
\end{IEEEeqnarray}
\fi
We will use this upper bound in the numerical evaluations reported in Section~\ref{sec:numerical_results}.
\end{rem}


\begin{rem}
  A converse bound that holds when the per-coherence-interval power constraint~\eqref{eq:avp} is replaced by the less stringent (and perhaps more common) per-codeword power constraint
  \begin{IEEEeqnarray}{rCL}
    \tr\lefto\{\herm{\matC_m}\matC_m\right\}\leq \tfdiv  \cohtime \snr
  \end{IEEEeqnarray}
  can be obtained by evaluating the supremum in~\eqref{eq:metaconverse_bound} and~\eqref{eq:information_spectrum} over all $\{\matSigma_k\}_{k=1}^{\tfdiv}$ that satisfy 
  \begin{IEEEeqnarray}{rCL}
    \sum_{k=1}^{\tfdiv} \tr\{\matSigma_k\}\leq \tfdiv \cohtime \snr.
  \end{IEEEeqnarray}
\end{rem}



\section{Bounds on the Coding Rate for Orthogonal Space-Time Codes} 
\label{sec:transmit_diversity_or_spatial_multiplexing}
In the previous section, we provided bounds on the maximum coding rate without imposing any constraint on how the multiple antennas available at the transmitter should be used. 
In this section, we focus on orthogonal space-time codes that use the available transmit antennas to provide transmit diversity and, hence, improve reliability. 
Specifically, we consider a setup where an outer code, defined along the same lines as in Definition~\ref{dfn:channel_code}, is combined with a specific orthogonal space-time inner code.
By treating this inner code as part of the channel, one can obtain achievability and converse bounds similar to the ones reported in Theorems~\ref{thm:dt_lb} and~\ref{thm:metaconverse_upper_bound}.
For simplicity, we shall focus on the $2\times 2$ and $4\times 4$ MIMO configurations.

\markblue{These bounds, which pertain to the case when the transmit antennas are used to provide full spatial diversity, are then compared in Section~\ref{sec:numerical_results} to the general bounds in Theorems~\ref{thm:dt_lb} and~\ref{thm:metaconverse_upper_bound}. This will allow us to characterize the rate penalty incurred by employing diversity-exploiting transmission strategies.
}

%
%

\subsection{$2\times 2$ Case: Alamouti} 
\label{sec:2times_2_mimo_}   
For the $2\times 2$ case, we consider an Alamouti  space-time inner code~\cite{alamouti98-10a}.
In order to analyze the finite-blocklength performance of such a scheme when CSI is not \emph{a priori} available at the receiver, we proceed as in Section~\ref{sec:bounds_on_the_maximal_coding_rate}: we first obtain a closed-form expression for the  output distribution induced by the Alamouti scheme, and then use this output distribution to obtain a DT lower bound and a MC upper bound on the maximum coding rate obtainable with such a scheme.

We assume that the coherence interval $\cohtime$ is even, and we let the $\cohtime\times 2$ input matrix $\matX_k$ in~\eqref{eq:channel_model} be given by
\begin{IEEEeqnarray}{rCL}\label{eq:alamouti_input}
  \matX_k=[\veca_k \quad \alshouffle(\veca_k)]
\end{IEEEeqnarray}
where $\veca_k$ is an $\cohtime$-dimensional vector satisfying $\vecnorm{\veca_k}^2=\snr\cohtime/2$, and
where the function $\alshouffle:\complexset^{\cohtime}\to \complexset^{\cohtime}$ maps an input vector $\veca$ into the output vector $\vecb$ according to the Alamouti rule~\cite{alamouti98-10a}:
\begin{IEEEeqnarray}{rCL}\label{eq:alamouti_shouffle}
  [\vecb]_{2l-1}&=&[\alshouffle(\veca)]_{2l-1}=\conj{[\veca]}_{2l}, \quad l=1,2,\dots,\cohtime/2 \IEEEyesnumber\IEEEyessubnumber\\
  {[\vecb]}_{2l}&=&[\alshouffle(\veca)]_{2l}=-\conj{[\veca]}_{2l-1}, \quad l=1,2,\dots,\cohtime/2\IEEEyessubnumber.  \IEEEeqnarraynumspace
\end{IEEEeqnarray}
In Lemma~\ref{lem:alamouti_pdf} below, we provide the pdf of the channel output $\rmatY$ induced by an input matrix $\rmatX$  constructed as in~\eqref{eq:alamouti_input} and whose first column $\rveca$ is uniformly distributed over the hypersphere of radius $\sqrt{\snr\cohtime/2}$ (this corresponds to USTM for the case of a single transmit antenna).
\begin{lem}\label{lem:alamouti_pdf}
  Assume that $\txant=\rxant=2$ and that $\cohtime$ is even and larger or equal to $4$. 
  Let
  \begin{IEEEeqnarray}{rCL}\label{eq:input_matrix_alamouti}
    \rmatX=[\rveca \quad \alshouffle(\rveca)]
  \end{IEEEeqnarray}
  where $\alshouffle(\cdot)$ is defined in~\eqref{eq:alamouti_shouffle} and where $\rveca=\sqrt{\snr\cohtime/2}\,\rvecu$, with $\rvecu$ being an isotropically distributed unit-norm $\cohtime$-dimensional complex random vector.
  Let $\rmatY=\bigl[ \rvecy_1\,\, \rvecy_2\bigr]=\rmatX\rmatH+\rmatW$, where $\rmatH\in \complexset^{2\times 2}$ and $\rmatW\in\complexset^{\cohtime\times 2}$ are defined similarly as in~\eqref{eq:channel_model}.
  Furthermore, let
  \begin{IEEEeqnarray}{rCL}
    \hat{\rmatY}=[\rvecy_1\quad \alshouffle(\rvecy_1)\quad \rvecy_2\quad  \alshouffle(\rvecy_2)]
  \end{IEEEeqnarray}
  and let $\Sigma_1$ and $\Sigma_3$ (with realizations $\sigma_1$ and $\sigma_3$, respectively) be the first and the third largest eigenvalue of the $4\times 4$ matrix $\snr\cohtime/(2+
  \snr\cohtime)\herm{\hat{\rmatY}}\hat{\rmatY}$.\footnote{The matrix $\herm{\hat{\rmatY}}\hat{\rmatY}$ has two distinct positive eigenvalues with multiplicity two almost surely.}
  Then, the pdf of $\rmatY$ is given by
  \begin{IEEEeqnarray}{rCL}\label{eq:out_dist_ala}   f_{\rmatY}(\matY)=\frac{\exp\lefto(\tr\lefto\{\herm{\matY}\matY\right\}\right)}{\pi^{2\cohtime}(1+\snr\cohtime/2)^{2\cohtime}}\frac{\Gamma(\cohtime)}{(\sigma_1-\sigma_3)^4}\det\lefto\{\matM(\sigma_1,\sigma_3)\right\}\IEEEeqnarraynumspace
  \end{IEEEeqnarray}
  where the $4\times 4$ matrix $\matM$ is given by
  \iftwocol
  \begin{IEEEeqnarray}{rCL}
    \mat
    \scriptstyle
    e^{\sigma_1}\altgamma(\cohtime-5,\sigma_1) & \scriptstyle (\cohtime-2)\sigma_1^{\cohtime-3} & \scriptstyle (\cohtime-3)\sigma_1^{\cohtime-4} & \scriptstyle (\cohtime-4)\sigma_1^{\cohtime-5}\\
     \scriptstyle e^{\sigma_1}\altgamma(\cohtime-4,\sigma_1) &\scriptstyle \sigma_1^{\cohtime-2} &\scriptstyle \sigma_1^{\cohtime-3} &\scriptstyle \sigma_1^{\cohtime-4}\\
    \scriptstyle e^{\sigma_3}\altgamma(\cohtime-5,\sigma_3) &\scriptstyle (\cohtime-2)\sigma_3^{\cohtime-3} &\scriptstyle (\cohtime-3)\sigma_3^{\cohtime-4} &\scriptstyle (\cohtime-4)\sigma_3^{\cohtime-5}\\
    \scriptstyle e^{\sigma_3}\altgamma(\cohtime-4,\sigma_3) &\scriptstyle \sigma_3^{\cohtime-2} &\scriptstyle \sigma_3^{\cohtime-3} &\scriptstyle \sigma_3^{\cohtime-4}\\
    \emat    \notag
  \end{IEEEeqnarray}
  \else
  \begin{IEEEeqnarray}{rCL}
    \matM=
    \mat
    
    e^{\sigma_1}\altgamma(\cohtime-5,\sigma_1) &  (\cohtime-2)\sigma_1^{\cohtime-3} &  (\cohtime-3)\sigma_1^{\cohtime-4} &  (\cohtime-4)\sigma_1^{\cohtime-5}\\
      e^{\sigma_1}\altgamma(\cohtime-4,\sigma_1) & \sigma_1^{\cohtime-2} & \sigma_1^{\cohtime-3} & \sigma_1^{\cohtime-4}\\
     e^{\sigma_3}\altgamma(\cohtime-5,\sigma_3) & (\cohtime-2)\sigma_3^{\cohtime-3} & (\cohtime-3)\sigma_3^{\cohtime-4} & (\cohtime-4)\sigma_3^{\cohtime-5}\\
     e^{\sigma_3}\altgamma(\cohtime-4,\sigma_3) & \sigma_3^{\cohtime-2} & \sigma_3^{\cohtime-3} & \sigma_3^{\cohtime-4}\\
    \emat    
  \end{IEEEeqnarray}
  \fi
  if $\cohtime>4$, and by 
  \begin{IEEEeqnarray}{rCL}
    \mat
    e^{\sigma_1} & 2\sigma_1 & 1 & 0 \\
    e^{\sigma_1} & \sigma_1^2 & \sigma_1 & 1 \\
    e^{\sigma_3} & 2\sigma_3 & 1 & 0 \\
    e^{\sigma_3} & \sigma_3^2 & \sigma_3 & 1 \\
    \emat
  \end{IEEEeqnarray}
  if $\cohtime=4$.
\end{lem}
\begin{IEEEproof}
  The proof follows along the same lines as the proof of Lemma~\ref{lem:ustm_output}.
\end{IEEEproof}
\begin{rem}
  Note that although $\rveca$ in~\eqref{eq:input_matrix_alamouti} is isotropically distributed, the matrix $\rmatX$ is not.
  Hence, $\rmatX$ does not follow a USTM distribution.
\end{rem}

Treating the Alamouti space-time inner code as part of the channel, we next report lower and upper bounds on the maximum coding rate $\Rmaxalap$ achievable when an Alamouti space-time inner code is used. 
These bounds rely on the closed-form expression for $f_{\rmatY}(\cdot)$ given in~\eqref{eq:out_dist_ala}.

\subsubsection{DT lower bound} 
\label{sec:ustm_dependence_testing_lower_bound_ala}
We provide first an achievability bound, which is based on the DT bound~\cite[Th.~22]{polyanskiy10-05a}.
\begin{thm}\label{thm:}
  Let $\{\rmatZ_k\}_{k=1}^{\tfdiv}$ be independent complex Gaussian $\cohtime\times 2$ matrices with \iid $\jpg(0,1)$ entries. 
  Let
  \begin{IEEEeqnarray}{rCL}
    \matD=\diag\biggl\{1+\frac{\snr\cohtime}{2},\, 1+\frac{\snr\cohtime}{2},\,\underbrace{1,\,\dots,\,1}_{\cohtime-2}\biggr\}
  \end{IEEEeqnarray}
  $\rmatV_k=\bigl[ \rvecv_{k,1} \,\, \rvecv_{k,2}\bigr]=\matD^{1/2}\rmatZ_k$, and
  \begin{IEEEeqnarray}{rCL}
    \hat{\rmatV}_k=\bigl[ \rvecv_{k,1}  \,\alshouffle(\rvecv_{k,1}) \, \rvecv_{k,2} \, \alshouffle(\rvecv_{k,2}) \bigr]
  \end{IEEEeqnarray}
  where the function $\alshouffle(\cdot)$ was defined in~\eqref{eq:alamouti_shouffle}.
  Furthermore, let $\Sigma_{k,1}$ and $\Sigma_{k,3}$ be the first and third largest eigenvalue of $(\snr\cohtime/(2+\snr\cohtime))\herm{\hat{\rmatV}_k}\hat{\rmatV}_k$ (which are positive and distinct almost surely), let
  \iftwocol
  \begin{IEEEeqnarray}{rCL}\label{eq:info_density_block_alamouti}    S_k&=&\tr\lefto\{\herm{\rmatZ_k}\matD\rmatZ_k\right\}-\tr\lefto\{\herm{\rmatZ_k}\rmatZ\right\}-\ln\Gamma(\cohtime) \notag\\
    &&+\ln\det\{\matM(\Sigma_{k,1},\Sigma_{k,3})\}-4\ln(\Sigma_{k,1}-\Sigma_{k,3}) \IEEEeqnarraynumspace
  \end{IEEEeqnarray}
  \else
  \begin{IEEEeqnarray}{rCL}\label{eq:info_density_block_alamouti}    S_k=\tr\lefto\{\herm{\rmatZ_k}\matD\rmatZ_k\right\}-\tr\lefto\{\herm{\rmatZ_k}\rmatZ\right\}-\ln\Gamma(\cohtime)+\ln\det\{\matM(\Sigma_{k,1},\Sigma_{k,3})\}-4\ln(\Sigma_{k,1}-\Sigma_{k,3}) \IEEEeqnarraynumspace
  \end{IEEEeqnarray}
  \fi
  and let
  \begin{IEEEeqnarray}{rCL}
      \epsilon\sub{ala}(\ncod)=   \Ex{}{\exp\lefto\{-\left[\sum_{k=1}^{\tfdiv}S_{k}-\ln(\ncod-1)\right]^{+}\right\}}. \IEEEeqnarraynumspace
  \end{IEEEeqnarray}
  Then
  \begin{IEEEeqnarray}{rCL}
    \Rmaxalap\geq \max \left\{ \frac{\ln \ncod}{\cohtime\tfdiv} \sothat  \epsilon\sub{ala}(\ncod)\leq \epsilon\right\}.
  \end{IEEEeqnarray}
\end{thm}
\begin{IEEEproof}
  The proof follows along the same lines as the proof of Theorem~\ref{thm:dt_lb}.
\end{IEEEproof}

\subsubsection{MC upper bound} 
\label{sec:metaconverse_upper_bound_ala}
Using~\cite[Th.~22]{polyanskiy10-05a} and~\cite[Eq.~(102)]{polyanskiy10-05a} we obtain the following converse bound on $\Rmaxalap$.
\begin{thm}\label{thm:alamouti_converse}
  Let $S_k$ be defined as in~\eqref{eq:info_density_block_alamouti}. Then
  \iftwocol
  \begin{multline}
    \Rmaxalap\\ \leq \inf_{\lambda>0} \frac{1}{\bl}\left[\lambda -\ln\lefto(\Pr\lefto\{\sum_{k=1}^{\tfdiv}S_k\leq \lambda\right\}-\epsilon\right)\right].
  \end{multline}
  \else
  \begin{IEEEeqnarray}{rCL}
    \Rmaxalap\leq \inf_{\lambda>0} \frac{1}{\bl}\left[\lambda -\ln\lefto(\Pr\lefto\{\sum_{k=1}^{\tfdiv}S_k\leq \lambda\right\}-\epsilon\right)\right].
  \end{IEEEeqnarray}
  \fi

\end{thm}
\begin{IEEEproof}
    The proof follows along the same lines as the proof of Theorem~\ref{thm:metaconverse_upper_bound}.
\end{IEEEproof}
%

\subsection{$4\times 4$ Case: Frequency-Switched Transmit Diversity} 
\label{sec:4times_4_mimo_}
  Since no generalization of the Alamouti space-time inner code exists beyond the $2\times 2$ configuration~\cite{tarokh99-07a}, we consider instead for the $4\times 4$ case the combination of Alamouti and  frequency-switched transmit diversity (FSTD) used in LTE~\cite[Sec.~11.2.2.1]{sesia11}. 
  According to this scheme, in the odd time-frequency slots only transmit antennas 1 and 2 are used, and in the even time-frequency slots only transmit antennas 3 and 4 are used. 
  In each time-frequency slot, an Alamouti space-time inner code is used for transmission.
  For example, for the case $\cohtime=4$, this scheme results in the following $4\times 4$ input matrix
  \begin{IEEEeqnarray}{rCL}
    \mat 
    a_1 & a_2 & 0  & 0\\
    0 & 0  & b_1 & b_2 \\
    -\conj{a_2} & \conj{a_1} & 0 & 0\\
    0 & 0 & -\conj{b_2} & \conj{b_1}
    \emat
  \end{IEEEeqnarray}
  where $\abs{a_1}^2+\abs{a_2}^2=\abs{b_1}^2+\abs{b_2}^2=\snr$.
  The combination of Alamouti and FSTD transforms a $4\times 4$ MIMO channel with coherence interval $\cohtime$ into two parallel $2\times 4$ MIMO channels with coherence interval $\cohtime/2$.
  Upper and lower bounds on the maximum coding rate achievable with this scheme can be obtained using a similar approach as for the $2\times 2$ case.


\section{Numerical Results} 
\label{sec:numerical_results}
Since a 5G standard for mission-critical MTC is not available yet, we base our numerical simulations on the setup analyzed in~\cite{lozano10-09a}. 
Specifically, we assume that packets of $n=168$ symbols are used to transmit within one millisecond $14$ OFDM symbols, each one consisting of $12$ tones. The available bandwidth is of $10$ MHz at a central frequency of $2$ GHz.
In a typical urban environment, one can obtain $12$ frequency-diversity branches by spreading the tones uniformly over the available bandwidth~\cite{lozano10-09a}.
Throughout, we set $\snr=6$~dB.
We  consider both the case where the packet error rate is $\epsilon=10^{-3}$, which may be appropriate for the exchange of short packets carrying control signaling,
and the case $\epsilon=10^{-5}$, which may be relevant for the transmission of critical information, e.g., in traffic-safety applications~\cite{metis-project-deliverable-d1.113-04a,popovski14-10a}.
For both cases, we shall compute the DT lower bound~\eqref{eq:dt_bound} and the MC upper bound~\eqref{eq:information_spectrum}.
The numerical evaluation of the MC upper bound~\eqref{eq:information_spectrum} is  challenging because it involves a maximization over the diagonal matrices $\{\matSigma_k\}$.
 As for the outage capacity in~\eqref{eq:outage_capacity}, the symmetry in~\eqref{eq:information_spectrum} suggests  that the supremum over $\{\matSigma_k\}_{k=1}^{\tfdiv}$ is achieved when
  \begin{IEEEeqnarray}{rCL}\label{eq:Telatar_at_FBL}
    \matSigma_k=\frac{\snr}{m_k}\diag\{\underbrace{1,\dots,1}_{m_k},\underbrace{0,\dots,0}_{\txant-m_k}\}
  \end{IEEEeqnarray}
  for some $m_k\in \{1,\dots,\txant\}$, $k\in \{1,\dots,\tfdiv\}$. We can think of~\eqref{eq:Telatar_at_FBL} as a finite-blocklength equivalent of Telatar conjecture~\cite{telatar99-11a}.
  Although far from conclusive, the numerical results reported in this section support the validity of this conjecture.
\paragraph*{Control signaling} 
\label{par:control_information_in_lte_}
\begin{figure}[t]
  \centering
    \includegraphics[width=\figwidth]{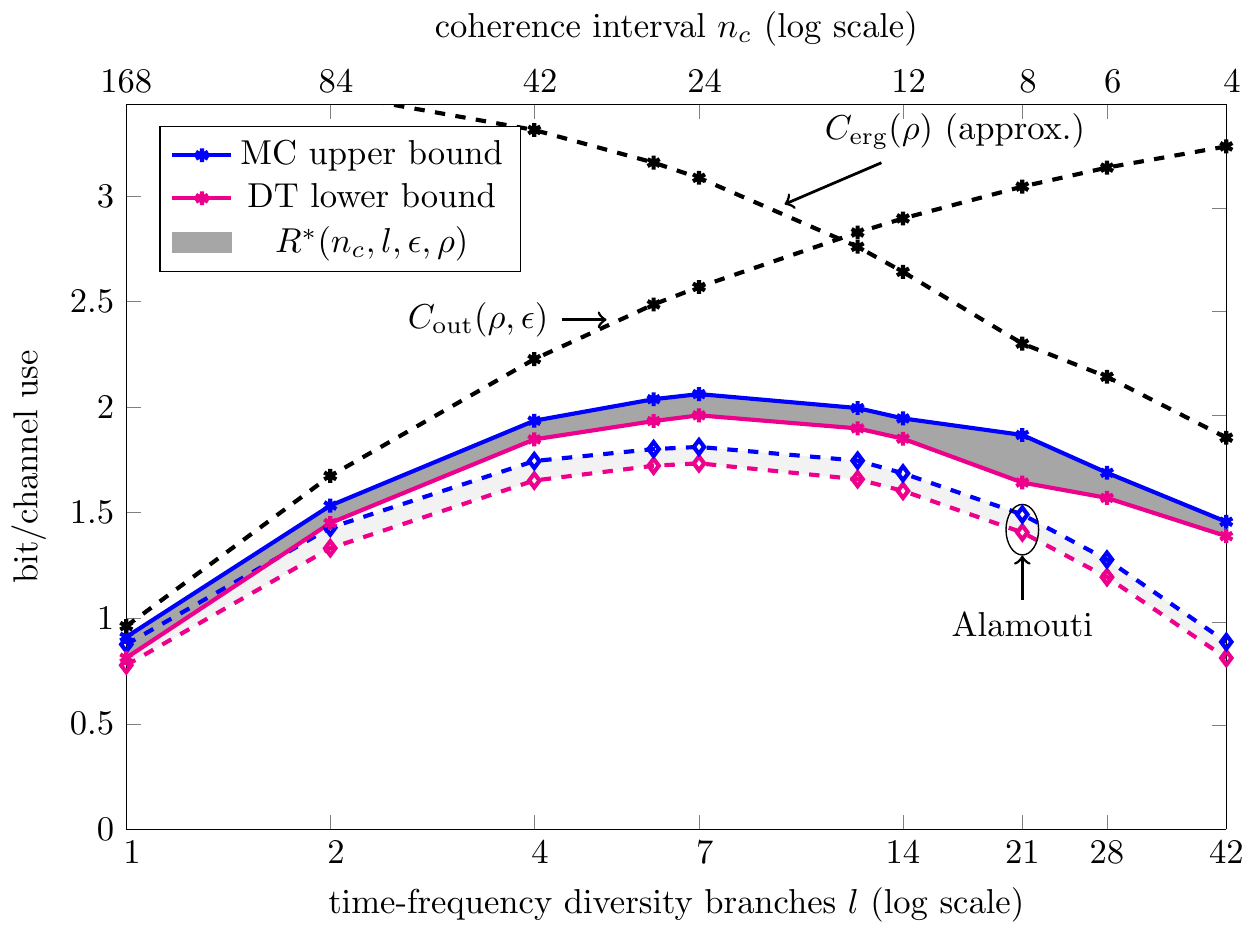}
  \caption{$\txant=\rxant=2$, $n=168$, $\epsilon=10^{-3}$, $\snr=6\dB$.
  Because of computational complexity, in the MC upper bound~\eqref{eq:information_spectrum} the supremum over $\{\matSigma_k\}_{k=1}^\tfdiv$ is restricted to  $\{\matSigma_k\}_{k=1}^\tfdiv$ of the form given in~\eqref{eq:Telatar_at_FBL} when $\tfdiv>7$.}
  \label{fig:figs_2x2_snr_6dB_eps_0.001_snr6eps03M2}
\end{figure}
In Fig.~\ref{fig:figs_2x2_snr_6dB_eps_0.001_snr6eps03M2} we plot\footnote{The numerical routines used to obtain these results are available at \url{https://github.com/yp-mit/spectre}} the DT lower bound~\eqref{eq:dt_bound} and the MC upper bound~\eqref{eq:information_spectrum} for the $2\times 2$ case. Here, $\epsilon=10^{-3}$.
These bounds delimit $\Rmaxp$ tightly and demonstrate that $\Rmaxp$ is not monotonic in the coherence interval $\cohtime$, but that there exists an optimal value $\cohtime^*$, or, equivalently, an optimal number $\tfdiv^*=\bl/\cohtime^*$ of time-frequency diversity branches, that maximizes $\Rmaxp$. 
A similar observation was reported in~\cite{yang12-09a} for the single-antenna case.
For $\cohtime<\cohtime^*$, the cost of estimating the channel dominates. 
For $\cohtime>\cohtime^*$, the bottleneck is the limited number of time-frequency diversity branches offered by the channel.  
For the parameters considered in Fig.~\ref{fig:figs_2x2_snr_6dB_eps_0.001_snr6eps03M2}, the optimal coherence interval length is $\cohtime^*\approx 24$, which corresponds to about~$7$ time-frequency diversity branches.

In the figure, we also plot the outage capacity $\Cout(\snr,\epsilon)$ in~\eqref{eq:outage_capacity} as a function of the number of time-frequency diversity branches $\tfdiv=\bl/\cohtime$ (with $\bl=168$), and a lower bound on the ergodic capacity $\Cerg(\snr)$ as a function of the coherence interval $\cohtime$.
This lower bound on $\Cerg(\snr)$, which is obtained by computing the mutual information on the RHS of~\eqref{eq:erg_cap} for the case when $\rmatX$ is USTM-distributed and by optimizing over the number of active transmit antennas, approximates $\Cerg(\snr)$ accurately already at moderate SNR values~\cite{devassy15-07a}.

As shown in the figure,  $\Cout(\epsilon,\snr)$ provides a good approximation for $\Rmaxp$ only when~$\tfdiv$ is small ($\cohtime\approx \bl$), i.e., when the fading channel is essentially constant over the duration of the packet (quasi-static scenario).
Furthermore, $\Cout(\epsilon,\snr)$ fails to capture the loss in throughput due to the channel estimation overhead, which is relevant for small $\cohtime$. 
For example, for $\cohtime=4$, the outage capacity overestimates~$\Rmaxp$ by a factor two.

The lower bound on  $\Cerg(\snr)$ plotted in the figure approximates~$\Rmaxp$ poorly when~$\cohtime$ is large.
For example, it overestimates $\Rmaxp$ by a factor four when $\cohtime=168$.
As expected, the approximation gets better as $\cohtime$ becomes smaller.

The number of active transmit antennas $\txantalt$ that maximizes the DT achievability bound is $\txantalt=2$  (both antennas  active) for  $1\leq \tfdiv \leq 21$, and it is $\txantalt=1$ (only one antenna  active) for $\tfdiv>21$.
The lower bound on $\Cerg(\snr)$, which also involves a maximization over the number of active antennas, exhibits the same behavior.  
We also note that the intersection between $\Cout(\epsilon,\snr)$ and $\Cerg(\snr)$ predicts coarsely the optimal number $\tfdiv^*$ of time-frequency diversity branches.


The optimal $\txantalt$ value for the MC upper bound~\eqref{eq:information_spectrum} is again $\txantalt=2$ for $1\leq \tfdiv \leq 21$ and $\txantalt=1$ for $\tfdiv > 21$.
Furthermore, the optimal\footnote{Because of computational complexity, for $\tfdiv>7$ only $\{\matSigma_k\}_{k=1}^\tfdiv$ of the form given in~\eqref{eq:Telatar_at_FBL} are considered.} $\{\matSigma_k\}_{k=1}^\tfdiv$ take all the same value and are equal to a $2\times 2$ scaled identity matrix for $1\leq \tfdiv\leq 14$ and for $\tfdiv=28$, and to a $2\times 2$ diagonal matrix with diagonal entries equal to $\snr$ and to $0$, respectively, for $\tfdiv=21$ and $\tfdiv=42$.


  In the same figure, we plot the achievability and the converse bounds for the case when an Alamouti code is used as inner code.
One can see that for small values of $\tfdiv$, the Alamouti scheme is almost optimal, but the gap between the DT lower bound and the Alamouti converse increases as $l$ grows. 
This is in agreement with the findings based on an outage-capacity analysis reported in~\cite{lozano10-09a}.
However, in contrast to what has been observed for the outage capacity, our bounds on $\Rmaxp$ reveal that it is better to switch off the second transmit antenna when $l$ is large. 
In this regime, the cost of estimating the channel resulting from the use of a second antenna overcomes the advantage of having additional spatial degrees of freedom.

We would like to emphasize that, in contrast to our approach, outage-capacity-based analyses are inherently insensitive to the cost of estimating the fading parameters and are therefore not suitable to capture the channel-estimation overhead.
Although the high-SNR ergodic capacity approximation~\eqref{eq:erg_high_snr} and the DMT for the case of no \emph{a priori} CSI~\eqref{eq:dmt_noncoherent} do predict that transmit antennas must be progressively switched off as $\tfdiv$ grows large, their predictions are coarse. 
Indeed, our numerical results suggest that the second transmit antenna should be switched off when $\tfdiv> 21$, or equivalently $\cohtime < 8$, whereas both~\eqref{eq:erg_high_snr} and~\eqref{eq:dmt_noncoherent} suggest that  the second antenna should be switched off only when $\cohtime\leq 3$.
Using both antennas when $3< \cohtime<8$  results in a rate loss that can be as large as $30\%$.

The gap between the DT lower bound and the MC upper bound in Fig.~\ref{fig:figs_2x2_snr_6dB_eps_0.001_snr6eps03M2} is largest around the value of $\cohtime$ (or equivalently $\tfdiv=\bl/\cohtime$) for which the second transmit antenna must be switched off. 
One could tighten both the DT and the MC bound by considering a larger class of input distributions (and the induced class of output distributions for the MC bound).
For example, one could drop the assumption that the input distribution is identical across coherence intervals. 
Indeed, using a different number of transmit antennas in different coherence intervals could be beneficial since it would essentially allow  one to extend the optimization in both~\eqref{eq:dt_bound_error_prob} and~\eqref{eq:information_spectrum} over fractional values of $\txantalt$.


%
%
%
\begin{figure}[t]
  \centering
    \includegraphics[width=\figwidth]{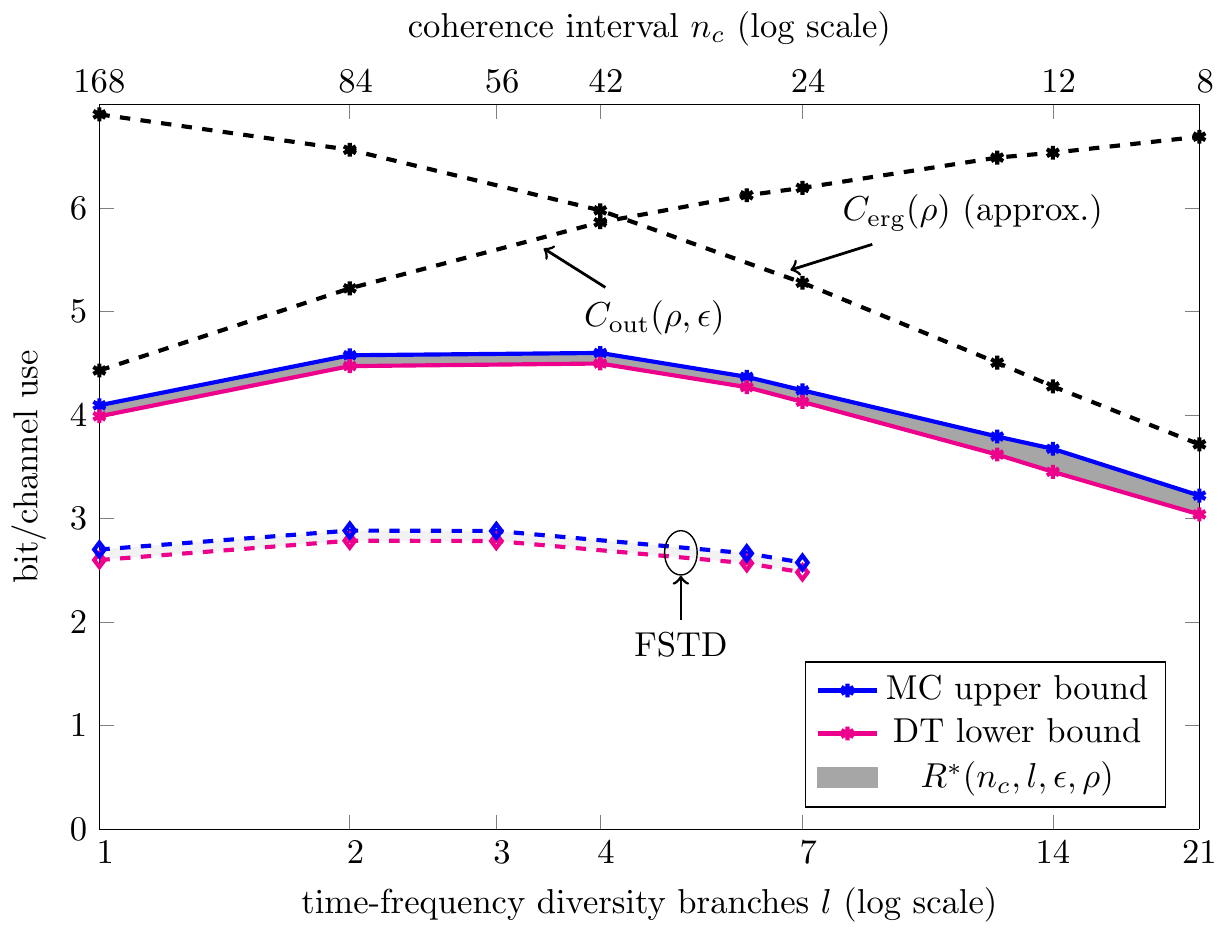}
    \caption{$\txant=\rxant=4$, $n=168$, $\epsilon=10^{-3}$, $\rho=6\dB$.
    Because of computational complexity, in the MC upper bound~\eqref{eq:information_spectrum} the supremum over $\{\matSigma_k\}_{k=1}^\tfdiv$ is restricted to  $\{\matSigma_k\}_{k=1}^\tfdiv$ of the form given in~\eqref{eq:Telatar_at_FBL}.}
  \label{fig:figs_4x4_snr_6dB_eps_1e-3_snr6eps03M4}
\end{figure}
In Fig.~\ref{fig:figs_4x4_snr_6dB_eps_1e-3_snr6eps03M4}, we present a similar comparison for the case of a $4\times 4$  system.
As shown in the figure, the gap between the MC upper bound and the DT lower bound is small, allowing for an accurate characterization of $\Rmaxp$. 
In contrast, the gap between the DT lower bound and the FSTD upper bound is large, which suggests that using all $4$ transmit antennas to provide spatial diversity is suboptimal even when the number of time-frequency diversity branches is limited (i.e., $\tfdiv$ is small).
As in the $2\times 2$ case, the transmit antennas should progressively be switched off as $\tfdiv$ increases, in order to mitigate the channel-estimation overhead. 
Specifically, the DT achievability bound is maximized by using $4$ transmit antennas ($\txantalt=4$) when $1\leq \tfdiv <
12$, by using $3$ antennas  when $\tfdiv=12$, and by using only two  antennas when $12< \tfdiv \leq 21$.
Also in this case, the lower bound on $\Cerg(\snr)$ and the MC upper bound exhibit a similar behavior. 
%

\begin{figure}[t]
  \centering
    \includegraphics[width=\figwidth]{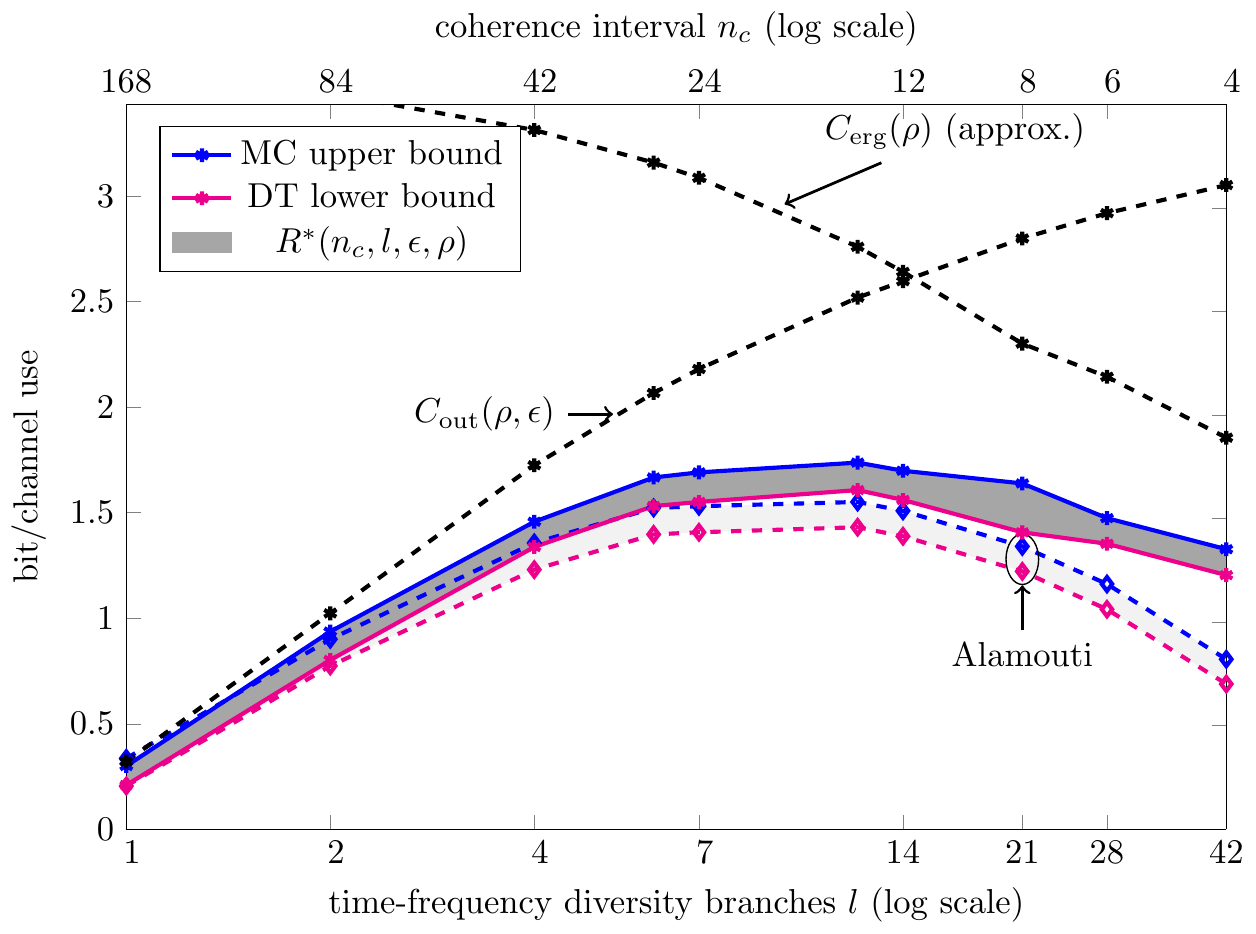}
  \caption{$\txant=\rxant=2$, $n=168$, $\epsilon=10^{-5}$, $\rho=6\dB$.
 Because of computational complexity, in the MC upper bound~\eqref{eq:information_spectrum} the supremum over $\{\matSigma_k\}_{k=1}^\tfdiv$ is restricted to  $\{\matSigma_k\}_{k=1}^\tfdiv$ of the form given in~\eqref{eq:Telatar_at_FBL} when $\tfdiv>7$.}
  \label{fig:figs_2x2_snr_6dB_eps_1e-5_snr6eps05M2}
\end{figure}
\begin{figure}[t]
  \centering
    \includegraphics[width=\figwidth]{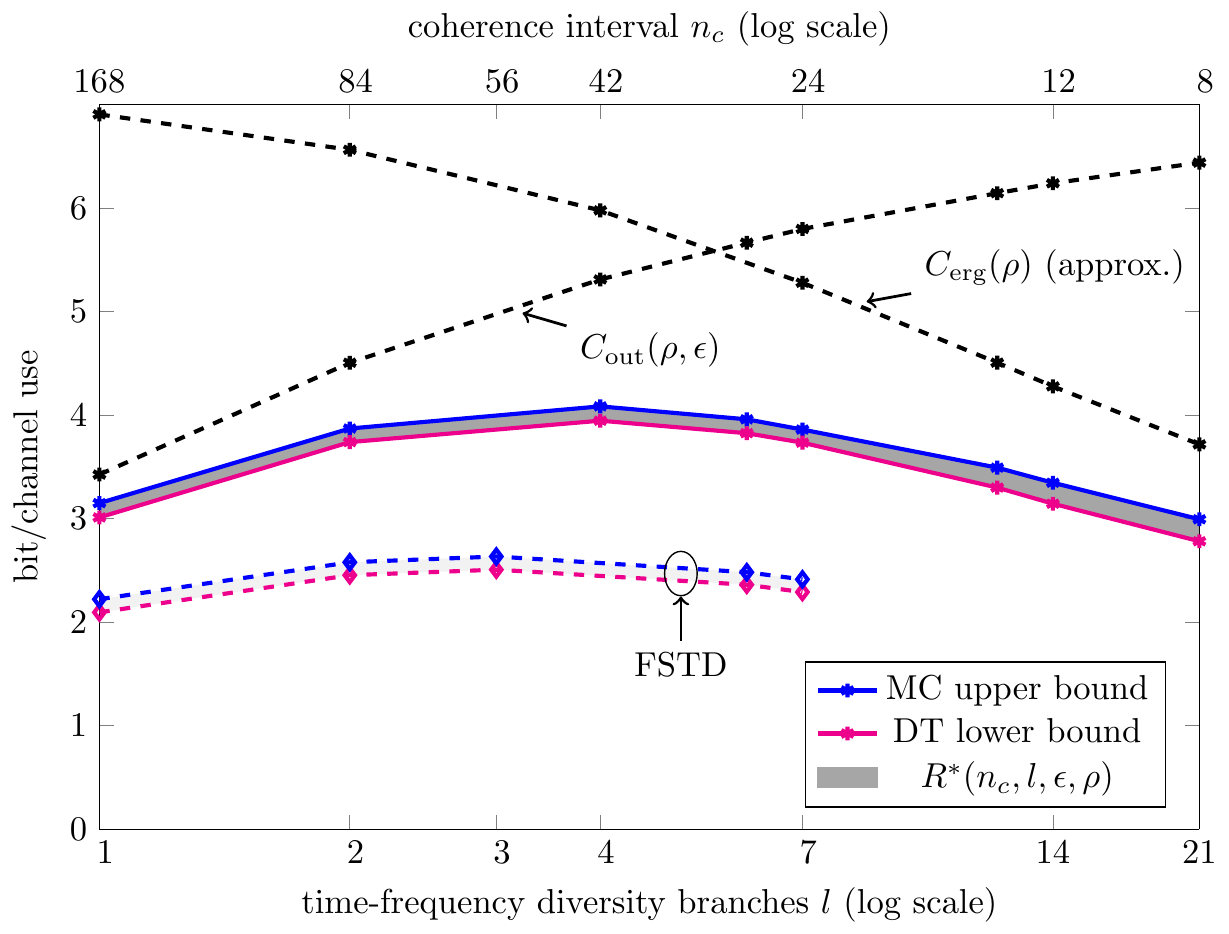}
    \caption{$\txant=\rxant=4$, $n=168$, $\epsilon=10^{-5},\rho=6\dB$.
    Because of computational complexity, in the MC upper bound~\eqref{eq:information_spectrum} the supremum over $\{\matSigma_k\}_{k=1}^\tfdiv$ is performed  only over  $\{\matSigma_k\}_{k=1}^\tfdiv$ values of the form given in~\eqref{eq:Telatar_at_FBL}.}
  \label{fig:figs_4x4_snr_6dB_eps_1e-5_snr6eps05M4}
\end{figure}

\paragraph*{Ultra-reliable communication} 
\label{par:ultra_highly_reliable_communication}
In Figs.~\ref{fig:figs_2x2_snr_6dB_eps_1e-5_snr6eps05M2} and~\ref{fig:figs_4x4_snr_6dB_eps_1e-5_snr6eps05M4}, we consider the case $\epsilon=10^{-5}$. 
We observe a similar behavior as for the case $\epsilon=10^{-3}$, with the difference that the gap between the optimal schemes and the orthogonal space-time schemes (Alamouti for the $2\times 2$ configuration, and FSTD for the $4\times 4$ case) becomes smaller. 
This comes as no surprise, since the higher reliability requirement makes the exploitation of transmit diversity advantageous.
  
\section{Conclusions} 
\label{sec:conclusions}
We presented finite-blocklength bounds on the maximum coding rate achievable over a MIMO Rayleigh block-fading channel, under the assumption that neither the transmitter nor the receiver have \emph{a priori} CSI.
Our bounds are explicit in the packet error rate $\epsilon$, the coherence interval $\cohtime$, and the number of time-frequency diversity branches $\tfdiv$.
Furthermore, they allow one to determine, for a fixed packet size $\bl=\cohtime\tfdiv$, the  number of time-frequency diversity branches and the  number of transmit antennas that maximize the rate. 
The optimal choice balances the rate gain resulting from exploiting of the available time-frequency-spatial resources, against the  cost of estimating the channel coefficients over these resources. 
The bounds  provide also an indication of whether the available transmit antennas should be used to provide transmit diversity or spatial multiplexing.

Our numerical results demonstrate that traditional infinite-blocklength performance metrics, such as the outage and the ergodic capacity, provide inaccurate estimates on the maximum coding rate when the packet size is short.
They further fail to capture the fundamental tradeoff between reliability, throughput, latency and channel-estimation overhead.
This suggests that the optimal design of the novel  low-latency, ultra-reliable MTC that will be provided by next-generation wireless systems must rely on a more refined analysis of the interplay between packet-error probability, communication rate, and packet size, than the one offered by traditional infinite-blocklength  performance metrics.

\appendices

\section{Proof of Theorem~\ref{thm:dt_lb}} 
\label{sec:proof_of_theorem_dt}
The transmitter uses only $\txantalt$ out of the available $\txant$ antennas. 
This yields an $\txantalt\times\rxant$ MIMO Rayleigh block-fading channel.
Let $\rmatX_k=\sqrt{\snr\cohtime/\txantalt}\rmatU_k$, $k=1,\dots,\tfdiv$, where $\{\rmatU_k\}_{k=1}^{\tfdiv}$ are independent, isotropically distributed $\cohtime\times\txantalt$  random matrices with orthonormal columns.
The induced channel outputs $\rmatY_k=\sqrt{\snr\cohtime/\txantalt}\rmatU_k\rmatH_k+\rmatW_k$, $k=1,\dots,\tfdiv$, are \iid $f_{\rmatY}$-distributed, where $f_{\rmatY}$ is given in~\eqref{eq:USTM_output_distribution}.
Let $\matU^{\tfdiv}=[\matU_1,\dots, \matU_{\tfdiv}]$.
Since the channel is block-memoryless, the information density~\cite[Eq.~(4)]{polyanskiy10-05a} can be decomposed as
\begin{IEEEeqnarray}{rCL}\label{eq:inf_density}
  \infoden{\matU^{\tfdiv}}{\matY^{\tfdiv}}=\sum_{k=1}^{\tfdiv} \infoden{\matU_k}{\matY_k}=\sum_{k=1}^{\tfdiv} \ln\frac{f_{\rmatY\given \rmatU}(\matY_k\given \matU_k)}{f_{\rmatY}(\matY_k)}
\end{IEEEeqnarray}
where
\begin{IEEEeqnarray}{rCL}
  f_{\rmatY\given \rmatU}
  (\matY_k\given\matU_k)
=
\frac{e^{-\tr\left\{\herm{\matY_k}\left(\matI_{\cohtime}+(\snr\cohtime/\txantalt)\matU_k\herm{\matU_k}\right)^{-1}\matY_k\right\}}}
{\pi^{\rxant\cohtime}(1+\snr\cohtime/\txantalt)^{\txantalt\rxant}}.\IEEEeqnarraynumspace
\end{IEEEeqnarray}
We next note that, for every $\cohtime\times\cohtime$ unitary matrix $\matV$,
\begin{IEEEeqnarray}{rCL}\label{eq:channel_law_invariance}
  f_{\rmatY\given\rmatU}(\matY\given \herm{\matV}\matU)=f_{\rmatY\given\rmatU}(\matV\matY\given \matU)
\end{IEEEeqnarray}
and
\begin{IEEEeqnarray}{rCL}\label{eq:output_law_invariance}
  f_{\rmatY}(\matV\matY)=f_{\rmatY}(\matY).
\end{IEEEeqnarray}
Consequently, the probability law of the information density $\infoden{\matU_k}{\rmatY_k}$ in~\eqref{eq:inf_density} (where $\rmatY_k\distas f_{\rmatY}$) does not depend on $\matU_k$. 
Without loss of generality, we shall then set $\matU_k=\bar{\matU}$, $k=1,\dots,\tfdiv$, with
\begin{IEEEeqnarray}{rCL}
  \bar{\matU}=
  \mat 
    \matI_{\txantalt}\\ 
    \veczero_{\cohtime-\txantalt\times \txantalt}
  \emat.
\end{IEEEeqnarray}
Using~\cite[Th.~22]{polyanskiy10-05a}, we conclude that there exists an $(\tfdiv,\cohtime,\ncod,\epsilon,\snr)$ code satisfying
\begin{IEEEeqnarray}{rCL}\label{eq:dt_bound_error_prob}
  \epsilon\leq \Ex{}{\exp\left\{ -\left[\sum_{k=1}^{\tfdiv} \infoden{\bar{\matU}}{\rmatY_k}-\ln(\ncod-1)\right]^{+} \right\}}\IEEEeqnarraynumspace
\end{IEEEeqnarray}
where the expectation is with respect to $\rmatY_k\distas f_{\rmatY\given \rmatU}(\cdot\given \bar{\matU})$.
Through algebraic manipulations, one can show that $\infoden{\bar{\matU}}{\rmatY_k}$ has the same distribution as the random variable $S_{k,\txantalt}$ in~\eqref{eq:S}.
Minimizing~\eqref{eq:dt_bound_error_prob} over the number of effectively used transmit antennas $\txantalt$, and solving the resulting inequality for the rate $(\ln \ncod)/(\cohtime\tfdiv)$ yields~\eqref{eq:dt_bound}.


\section{Proof of Theorem~\ref{thm:metaconverse_upper_bound}} 
\label{sec:proof_of_theorem_MC}
  Fix $1\leq \txantalt\leq \txant$.
  To upper-bound $\Rmaxp$, we use the meta-converse theorem for maximal error probability~\cite[Th.~31]{polyanskiy10-05a} with  auxiliary pdf
  \begin{IEEEeqnarray}{rCL}
    q_{\rmatY^\tfdiv}(\matY^\tfdiv)=\prod_{k=1}^{\tfdiv} f_{\rmatY}(\matY_k)
  \end{IEEEeqnarray}
  where $f_{\rmatY}$ is the USTM-induced output pdf defined in~\eqref{eq:USTM_output_distribution}.
  This yields
  \begin{IEEEeqnarray}{rCL}\label{eq:upper_bound_explicit_in_beta}
    \Rmaxp\leq \sup_{\matX^\tfdiv}\frac{1}{\bl}\ln\frac{1}{\displaystyle \beta_{1-\epsilon}\lefto(\matX^\tfdiv,q_{\rmatY^\tfdiv}\right)}
  \end{IEEEeqnarray}
  where the supremum is over all codewords $\matX^\tfdiv \in \complexset^{\cohtime \times \txant\tfdiv}$ satisfying the power constraint~\eqref{eq:subavp}, and where $\beta_{1-\epsilon}(\cdot,\cdot)$ is defined as in~\cite[Eq.~(105)]{polyanskiy10-05a}.\footnote{To be precise, the second argument of $\beta_{1-\epsilon}(\cdot,\cdot)$ in~\cite[Eq.~(105)]{polyanskiy10-05a} is an arbitrary probability measure. In our case, since the chosen probability measure is absolutely continuous, it is convenient to let the second argument of $\beta_{1-\epsilon}(\cdot,\cdot)$ be a pdf.}
By the Neyman-Pearson lemma, we have that
\begin{IEEEeqnarray}{rCL}
  \beta_{1-\epsilon}(\matX^{\tfdiv},q_{\rmatY^{\tfdiv}})=\Pr\lefto\{ \infoden{\matX^{\tfdiv}}{\rmatY^{\tfdiv}} \geq \gamma\right\}, \quad \rmatY^{\tfdiv}\distas q_{\rmatY^\tfdiv}
\end{IEEEeqnarray}
where $\gamma$ is the solution of
\begin{IEEEeqnarray}{rCL}
  \Pr\lefto\{ \infoden{\matX^{\tfdiv}}{\rmatY^{\tfdiv}} \leq \gamma\right\}=\epsilon, \quad \rmatY^{\tfdiv}\distas f_{\rmatY^\tfdiv\given \rmatX^\tfdiv}(\cdot \given \matX^\tfdiv)
\end{IEEEeqnarray}
and where $\infoden{\cdot}{\cdot}$ is defined as in~\eqref{eq:inf_density}.

For a given codeword $\matX^{\tfdiv}=[\matX_1,\dots,\matX_{\tfdiv}]$, let 
\begin{IEEEeqnarray}{rCL}
  \matX_k\herm{\matX_k}=\matV_k\matSigma_k\herm{\matV_k}, \quad k=1,\dots,\tfdiv.
\end{IEEEeqnarray}
Here, $\matV_k \in \complexset^{\cohtime\times\txant}$ contains the eigenvectors of $\matX_k\herm{\matX_k}$, and $\matSigma_k\in\complexset^{\txant\times\txant}$ is a diagonal matrix with nonnegative entries containing the $\txant$ eigenvalues of $\herm{\matX_k}\matX_k$.
It follows from~\eqref{eq:channel_law_invariance} and~\eqref{eq:output_law_invariance} that $\beta_{1-\epsilon}(\matX^{\tfdiv},q_{\rmatY^{\tfdiv}})$ depends on $\matX^{\tfdiv}$ only through the diagonal matrices $\{\matSigma_k\}_{k=1}^\tfdiv$.
Hence, we can replace the infimum over $\matX^{\tfdiv}$ in~\eqref{eq:upper_bound_explicit_in_beta} by an infimum over $\{\matSigma_k\}_{k=1}^{\tfdiv}$.

We continue the proof by noting that, when $\rmatY^{\tfdiv}\distas f_{\rmatY^\tfdiv\given \rmatX^\tfdiv}(\cdot \given \matX^\tfdiv)$,  the information density $\infoden{\matX^{\tfdiv}}{\rmatY^{\tfdiv}}$ is distributed as $\sum_{k=1}^{\tfdiv}\bar{S}_{k,\txantalt}$, with $\bar{S}_{k,\txantalt}$ defined in~\eqref{eq:P_inf}; and when $\rmatY^{\tfdiv}\distas q_{\rmatY^\tfdiv}$ the information density is distributed as $\sum_{k=1}^{\tfdiv}T_{k,\txantalt}$, with $T_{k,\txantalt}$ defined in~\eqref{eq:Q_inf}.
Finally,~\eqref{eq:metaconverse_bound} follows by minimizing over $\txantalt \in \{1,\dots,\txant\}$.


\bibliographystyle{IEEEtran}
\bibliography{IEEEabrv,publishers,confs-jrnls,giubib}

\end{document}